\documentclass[useAMS,usenatbib]{mn2e}
 \usepackage{lscape}
\usepackage{natbib}
\usepackage{epsfig}
\usepackage{rotating}
\usepackage{amsmath}
\usepackage{tabu}

\def\lesssim{\,\lower2truept\hbox{${<\atop\hbox{\raise4truept\hbox{$\sim$}}}$}\,}
\def\gtrsim{\,\lower2truept\hbox{${>\atop\hbox{\raise4truept\hbox{$\sim$}}}$}\,}

\title[Multi-frequency polarimetry of a complete sample]{Multi-frequency polarimetry of a complete sample of PACO radio sources}
\author[V. Galluzzi et al.]{\parbox[t]{\textwidth}{
V.~Galluzzi$^{1,2}$\thanks{E-mail: vgalluzzi@unibo.it (VG)}, M.~Massardi$^{1}$, A.~Bonaldi$^{3,4}$, V.~Casasola$^{5}$, L.~Gregorini$^{1}$, T.~Trombetti$^{6,7,8}$, C.~Burigana$^{6,7,8}$, G.~De Zotti$^{9,10}$
, R.~Ricci$^{1}$, J.~Stevens$^{11}$, R.~D.~Ekers$^{12,13}$, L.~Bonavera$^{14}$, S.~di Serego Alighieri$^{5}$, E.~Liuzzo$^{1}$, M.~L\'opez-Caniego$^{15}$, A.~Mignano$^{1}$,
R.~Paladino$^{1}$, L.~Toffolatti$^{14,6}$ and M.~Tucci$^{16}$.
}
\vspace*{8pt}\\
$^{1}$INAF, Istituto di Radioastronomia, Via Piero Gobetti 101, I-40129 Bologna, Italy\\
$^{2}$Dipartimento di Fisica e Astronomia, Universit\`a di Bologna, via Ranzani 1, I-40126 Bologna, Italy\\
$^{3}$Jodrell Bank Centre for Astrophysics School of Physics \& Astronomy, The University of Manchester, Manchester M13 9PL, UK\\
$^{4}$SKA Organization, Lower Withington Macclesfield, Cheshire SK11 9DL, UK\\
$^{5}$INAF - Osservatorio Astrofisico di Arcetri, Largo Enrico Fermi 5, I-50125 Firenze, Italy\\ 
$^{6}$INAF-IASF Bologna, Via Piero Gobetti 101, I-40129 Bologna, Italy\\
$^{7}$Dipartimento di Fisica e Scienze della Terra, Universit\`a degli Studi di Ferrara, Via Giuseppe Saragat 1, I-44100 Ferrara, Italy\\
$^{8}$INFN-Sezione di Bologna, Via Irnerio 46, I-40126 Bologna, Italy\\
$^{9}$INAF, Osservatorio Astronomico di Padova, Vicolo dell'Osservatorio 5, I-35122 Padova, Italy\\
$^{10}$SISSA, via Bonomea 265, I-34136 Trieste, Italy\\
$^{11}$CSIRO Astronomy and Space Science, PO Box 76, Epping, NSW 1710, Australia\\
$^{12}$International Centre for Radio Astronomy Research, Curtin University, Bentley, WA 6102, Australia\\\
$^{13}$ARC Centre of Excellence for All-sky Astrophysics (CAASTRO), Redfern, NSW 2016, Australia\\
$^{14}$Departamento de F\'\i{sica} Universidad de Oviedo c. Calvo Sotelo s/n 33007 - OVIEDO (Espa\~na)\\
$^{15}$European Space Agency, ESAC, Camino bajo del Castillo, s/n, Urbanizaci\'{o}n Villafranca del Castillo,\\
\;\;\,Villanueva de la Ca\~{n}ada, Madrid, Spain\\
$^{16}$D\'epartement de Physique Th\'eorique and Center for Astroparticle Physics (CAP), University of Geneva,\\\;\;\, 24 quai Ernest Ansermet, CH-1211 Geneva, Switzerland}
\begin{document}

\date{}

\pubyear{2010}

\maketitle

\label{firstpage}

\begin{abstract}
We present high sensitivity polarimetric observations ($\sigma_P \simeq 0.6\,$mJy) in 6 bands covering the $5.5-38\,$GHz range of a complete sample of 53 compact extragalactic radio sources brighter than $200\,$mJy at $20\,$GHz. The observations, carried out with the Australia Telescope Compact Array (ATCA), achieved a $91\%$ detection rate (at $5\sigma$).  Within this frequency range the spectra of about $95\%$ of sources are well fitted by double power laws, both in total intensity and in polarisation, but the spectral shapes are generally different in the two cases. Most sources were classified as either steep-- or peaked--spectrum but less than $50\%$ have the same classification in total and in polarised intensity. No significant trends of the polarisation degree with flux density or with frequency were found. The mean variability index in total intensity of steep-spectrum sources increases with frequency for a $4-5$ year lag, while no significant trend shows up for the other sources and for the 8 year lag. In polarisation, the variability index, that could be computed only for the $8$ year lag, is substantially higher than in total intensity and has no significant frequency dependence.
\end{abstract}

\vspace*{5cm}

\section{Introduction}

At high radio frequencies ($\nu > 10\,$GHz) the bright ($S_{20\,{\rm GHz}}\gtrsim 100\,$mJy) extragalactic radio population is dominated by blazars (BL Lacs and Flat-Spectrum Radio Quasars, FSRQs), interpreted as sources whose relativistic jets are directed very close to the line-of-sight (see, e.g., Blandford \& K\"onigl 1979). The jets are collimated by intense magnetic fields. The blazar radio emission is synchrotron radiation from a series of blobs along the jet. The observed flux densities are blue-shifted and greatly boosted by relativistic effects.

In this scenario, the flat radio spectra are interpreted as the combination of components self-absorbed up to different frequencies. The spectra steepen when the emission becomes optically thin. Young and compact objects are characterised by spectra peaking at high frequencies while older sources are more extended and have spectra steepening at lower frequencies. Occasionally the spectra may be dominated by a single flaring blob. Measurements of polarimetric properties of radio-loud Active Galactic Nuclei (AGN) at centimetric and millimetric wavelengths carry information on magnetic fields and on the plasma in the inner, unresolved regions of their relativistic jets.

Even if synchrotron radiation can be up to $\simeq 70-80\%$ polarized, the polarization degree of extragalactic sources is rarely observed to be higher than $\sim 10\%$; the median values are $\sim 2.5\%$ at $20\,$GHz (Massardi et al. 2013). This is the result of several depolarisation effects. The measured polarised flux is an average of the emission from sub-structures with different orientations of the magnetic field. Further depolarisation may be due to differential Faraday rotation that may be amplified by its frequency dependence within the bandwidth. As a consequence, polarimetric observations generally require high sensitivities and a detailed knowledge of instrumental properties to perform accurate calibration (see Massardi et al. 2016a), a delicate procedure at high frequencies. For these reasons polarimetric surveys of large source samples at $\nu \gtrsim 20\,$GHz have become possible only recently thanks to the advent of large bandwidth instruments (see Galluzzi et al. 2016 for an up-to-date summary of the available observations).

To set the stage we give here an overview of the main high frequency polarisation surveys in the table \ref{tab:polcom}, reporting the detection rates and median polarisation fractions found at the frequency $\simeq 20\,$GHz.
\begin{table*}
\label{tab:polcom}
\caption{Summary of some of the surveys in polarization available at high radio frequencies. The tabulated detection rates and polarisation fractions (see the columns labelled as ``det. rate'' and ``pol. fraction'', respectively) refer in each case to the frequency closest to $20\,$GHz, when available.}
   \begin{tabular}{lccccl}
\hline
\scriptsize{\textbf{References}}            &\scriptsize{\textbf{Freq. (GHz)}}&\scriptsize{\textbf{\# sources}}&\scriptsize{\textbf{det. rate}}&\scriptsize{\textbf{pol. fraction}}&\scriptsize{\textbf{Notes}}\\
\hline
\scriptsize{Ricci et al. (2004)}& \scriptsize{$18.5$} & \scriptsize{$250$}  & \scriptsize{$68\%$}& \scriptsize{$2.7\%$}\footnote{This median value is found considering the flat-spectrum sub-sample only.}& \scriptsize{complete sample with $S_{5\,\rm GHz} > 1\,$Jy} \\
\scriptsize{Sadler et al. (2006)}& \scriptsize{$20$}   & \scriptsize{$173$}  & \scriptsize{$75\%$}& \scriptsize{$2.3\%$} & \scriptsize{complete sample with $S_{20\,\rm GHz} > 0.1\,$Jy}\\
\scriptsize{Massardi et al. (2008) AT20G-BSS}& \scriptsize{$4.8,\,8.6,\,20$} & \scriptsize{$320$}  & \scriptsize{$67\%$} & \scriptsize{$2.5\%$} & \scriptsize{AT20G bright ($S_{20\,\rm GHz} > 0.5\,$Jy) sample}\\
\scriptsize{Murphy et al. (2010); Massardi et al. (2011)}\\ \scriptsize{AT20G}& \scriptsize{$4.8,\,8.6,\,20$} & \scriptsize{$5890$} & \scriptsize{$72.5\%$}\footnote{The analysis is limited to objects with $S_{20\,\rm GHz}> 250\,$mJy.} & \scriptsize{$2.7\%$} & \scriptsize{survey $93\%$ complete with $S_{20{\,\rm GHz}}>40\,$mJy}\\
\scriptsize{Sajina et al. (2011)}& \scriptsize{$4.86,\,8.46,\,22.46,\,43.34$} & \scriptsize{$159$} & \scriptsize{$56\%$} & \scriptsize{$3.82\%$} & \scriptsize{complete sample of AT20G objects with}\\ & & & & & \scriptsize{$S_{20\,\rm GHz}> 40\,$mJy in an equatorial field}\\ 
\scriptsize{Battye et al. (2011)}& \scriptsize{$8.4,\,22,\,43$} & \scriptsize{$203$} & \scriptsize{$83\%$} & \scriptsize{$2.0\%$} & \scriptsize{WMAP sources follow-up}\\
\scriptsize{Massardi et al. (2013)}& \scriptsize{$4.8,\,8.6,\,18$} & \scriptsize{$193$} & \scriptsize{$91.4\%$}\footnote{The detection rate is  $\simeq 94\%$ considering the sub-sample of point-like objects.} & \scriptsize{$2\%$}& \scriptsize{complete sample with $S_{20{\,\rm GHz}}>500\,$mJy}\\
\scriptsize{Agudo et al. 2014}& \scriptsize{$15,\,86,\,229$} & \scriptsize{$211$} & \scriptsize{$88\%$} \footnote{The detection rate is reported for the $86\,$GHz observations.} & \scriptsize{$3.2\%$; $4.6\%$}\footnote{These values refer to quasars and BL Lacs, respectively.} & \scriptsize{complete sample with $S_{86\,\rm GHz}> 0.9\,$Jy}\\
\hline
   \end{tabular} 
\end{table*}
These studies did not detect any statistically significant relationship between polarisation fraction and total flux density (but all covered a limited flux density range) or with frequency. On the contrary, various studies of samples mostly selected at $1.4\,$GHz and covering broader flux density ranges, reported an  increase of the polarisation degree with decreasing flux density and with increasing frequency (Mesa et al. 2002; Tucci et al. 2004; Taylor et al. 2007; Grant et al. 2010; Subrahmanyan et al. 2010). These results were however challenged by Hales et al. (2014).

On the whole, our understanding of the high radio frequency polarisation properties of extragalactic sources is still limited. This has motivated the new high-sensitivity ($\sigma_P \simeq 0.6\,$mJy, a factor $\simeq 2$ better than the AT20G sensitivity) and nearly simultaneous, multi-frequency polarimetric observations over a wide spectral range ($5.5-38\,$GHz) whose results are reported here. We observed with the ATCA the {\it Planck}-ATCA Coeval Observations (PACO) ``faint sample'' (Massardi et al. 2016b), a complete sample of AT20G sources with $S_{20\,\rm GHz}\ge 200\,$mJy. The instrumental capabilities were exploited to reconstruct the polarimetric spectral profile to unprecedented  detail.

High frequency observations are easily affected by spectral, detection and variability-related biases. The present work, owing to the completeness of the sample, the high sensitivity and the close-in-time multi-frequency characterisation, allows us to minimise all these effects. Thus, we provide a new statistical assessment of the polarimetric properties of high frequency extragalactic radio sources, analysing and describing their spectral and polarisation behaviour with frequency. Moreover, exploiting also the AT20G and PACO observations in total intensity, we investigate the source variability for $4-5$ and $8\,$yr time lags at several frequencies up to $38\,$GHz.

Multi-frequency and multi-epoch polarimetric observations of radio sources are useful cosmological tools for several reasons. Radio sources are the most relevant foreground contaminants of the Cosmic Microwave Background (CMB) polarisation maps (on scales of up to $\sim 30\,$arcmin) in the relatively clean $70-100\,$GHz frequency range (Massardi et al. 2016a). A proper characterisation of the radio source contribution to the power spectrum in polarisation is essential for a precise assessment of the lensing B-mode signal (Hanson et al. 2013, POLARBEAR: Ade et al. 2014). This, in turn, is essential for the detection and the characterisation of the power spectrum of the primordial B-modes associated with the stochastic background of gravitational waves, the most ambitious goal of current and future CMB projects (PRISM Collaboration 2014). Because of the broad variety of polarized emission spectra, extrapolations from low frequencies ($<20\,$GHz) are inadequate to model the radio source contribution in CMB polarisation maps (Huffenberger et al. 2015).

At the same time, radio sources are the privileged calibrators both of the polarized intensity and of the polarisation angle for CMB experiments. The systematic errors due to inaccuracies in the calibration of the polarisation angle are becoming the limiting factor for CMB polarization experiments (e.g. Kaufman et al. 2016). Unfortunately, the number of compact, bright, highly polarised and stable enough extragalactic sources, suitable for accurately calibrating CMB polarization maps are rare. One of the purposes of our polarisation survey was to identify good candidates.

The paper is organised as follows. In Section \ref{obs} we briefly present the PACO project, its main results and the polarimetric observations of the selected sub-sample held in September 2014 with ATCA. In Section \ref{sec:maths} we describe the data reduction. In Section \ref{dataanaly} we discuss the data analysis and the spectral behaviours in total intensity and polarisation. In Section \ref{variability} we address the variability properties for our sample. Finally, in Section \ref{discusseconcl} we draw our conclusions.

\section{Observations}
\label{obs}

The PACO project yielded observations of $464$ AT20G sources in $65$ epochs between July 2009 and August 2010 in the $5.5-39\,$GHz frequency range with ATCA, nearly simultaneously (within $10$ days) with the {\it Planck} observations (Massardi et al. 2016b). The main goal of the project was to characterize, in combination with \textit{Planck} data, the total intensity spectra and their variability over a wide frequency range (at least from $5\,$GHz to $217\,$GHz, but up to $857\,$GHz for some sources).

A multi-frequency observing run was dedicated to polarimetry in September 2014 with ATCA. The sample was extracted from the ``faint'' PACO sample ($S_{20 \rm{GHz}}>200\,$mJy) selecting sources in the Southern Ecliptic Pole region (ecliptic latitude $<-75\,\deg$), where the satellite scanning strategy implies a better sensitivity.

The $5$ extended sources in the sample (i.e. sources larger than the PACO resolution of $\simeq 1\,$arcsec) were left out as the techniques to extract flux densities used in the PACO project underestimate their flux densities. The final complete sample comprises $53$ compact sources with $S_{20 \rm{GHz}}>200\,$mJy in the AT20G catalogue.

The spectral setup consisted in $3$ sets of $2\times 2\,$GHz CABB (Compact Array Broadband Backend) bands centred at $5.5-9\,$GHz, $18-24\,$GHz and $33-38\,$GHz, similar to those for the PACO program. A total of $\simeq 12\,$hr were allocated in three slots of $\simeq 4\,$hr each (including overheads and calibration) dedicated to each of the three bands over a period of least three days. Each object was observed for $\simeq 3\,$min in two $1.5\,$min scans at different hour angles for each frequency. Weather conditions were very good during all the campaign.

The array configuration was H214, a hybrid (antenna are displayed also along the N-S direction) compact array with a nominal spatial resolution ranging from $36$ to $5\,$arcsec in the $5.5-38\,$GHz interval if we consider only the $5$ most packed antennas. The resolution goes down to $0.5-4\,$arcsec considering also the longest baselines with the sixth antenna, but with a lower sensitivity. Adopting a hybrid configuration results in a more homogeneous $uv$-plane coverage of the largest spatial scales in less time. This results in a better imaging with respect to a non-hybrid solution, at least for our sample of mostly point-like objects.

\section{Data reduction}
\label{sec:maths} 

The data reduction was done using the MIRIAD software package (Sault, Teuben \& Wright 1995) treating each frequency separately, as indicated in the ATCA User's Guide\footnote{www.narrabri.atnf.csiro.au/observing/users\_guide.}. While loading the data, we corrected for the time-dependent instrumental xy-phase variation, exploiting the known signal injected from a noise diode mounted in one of the feeds of each antenna.

The reference flux density calibrator (at all frequencies) was the source PKS1934-638, a bright (point-like) GigaHertz peaked-spectrum (GPS) radio galaxy, stable, unpolarised (at least below $30-40\,$GHz), whose model (see Reynolds 1994; Sault 2003; Partridge et al. 2016) is encoded into MIRIAD itself. It is the only known source with all these characteristics in the Southern sky.



Once the calibration tables were derived, all solutions were ingested in the code for flux density extraction. To better characterize the source spectra, we decided to split each $2\,$GHz-wide frequency band in sub-bands. Each of them was calibrated separately. For total intensity we split each band into $512\,$MHz-wide sub-bands, as done in the past analyses for the PACO project. For polarised flux densities we split bands in only $2$ sub-bands to limit the $\Delta\nu^{-1/2}$ degradation in sensitivity.

Flux densities were estimated via the MIRIAD task UVFLUX. Our sources are known to exhibit linear polarisation (up to $\sim 10\%$, Massardi et al. 2008, 2013), defined by the $Q$ and $U$ Stokes parameters. Observations of the circular polarisation of extragalactic radio sources have demonstrated that it is very low, generally below $0.1-0.2\%$, at least one order of magnitude lower than the linear polarisation (Rayner et al. 2000), hence generally undetectable for our sources, given the $0.6\,$mJy sensitivity. Hence, the rms $\sigma_V$ (of the retrieved $V$ Stokes parameter) is used as a noise estimator of the total intensity flux density $I$. The polarised emission, $P$, is given by the simple relation:
\begin{equation}
P=\sqrt{Q^2+U^2-\sigma^2}\,.
\label{equ:PolFluDen}
\end{equation}
where the $\sigma^2$ term removes the noise bias on $P$ under the assumption of equal noise values, $\sigma$, for the $Q$ and $U$ parameters (e.g. Wardle \& Kronberg 1974).\footnote{The error associated to the bias correction is negligible and will be omitted in next subsections.}

The polarisation angle $\phi$ and fraction $m$ (usually in terms of a percentage) are:

\begin{eqnarray}
\phi&=&\frac{1}{2}\arctan{\left(\frac{U}{Q}\right)},\\
m&=&100\cdot P/I.
\end{eqnarray}

\subsection{Error budget}
Assuming Gaussian noise, the error scales as $1/\sqrt{N}$, $N$ being the number of correlations at a given $\nu$. A suitable estimate of the total intensity error is the sum in quadrature of $\sigma_V$ with a systematic term, mainly accounting for the calibration uncertainty. Based on the past experience with PACO observations and on a comparison between flux densities obtained from different calibrators, the calibration error amounts to $\sim 2.5\%$ of the $I$ flux density, giving for $\sigma_I$, the global error on I:
\begin{equation}
\sigma^2_I=\sigma^2_V+(0.025\,I)^2.
\label{equ:ErrTotInt}
\end{equation}
The error, $\sigma_P$, on the polarised flux density, $P$, can be derived from the eq.~(\ref{equ:PolFluDen}). A conservative $\sim 10\%$ error in the polarisation calibration is adopted in this case. Then, we have:
\begin{equation}
\sigma^2_P=\frac{Q^2\sigma^2_Q+U^2\sigma^2_U}{Q^2+U^2}+(0.1\,P)^2 \,,
\label{equ:ErrPolFlu}
\end{equation}
$\sigma_{Q,U}$ being the rms errors on the Stokes parameters $Q$ and $U$, respectively. This estimate of the calibration error is consistent with the differences between flux densities obtained using two different calibrators for reducing our data.

From error propagation, the global error on the polarisation fraction is:
\begin{equation}
\sigma^2_m=\left(\frac{\sigma_I}{I}\right)^2+\left(\frac{\sigma_P}{P}\right)^2,
\end{equation}
and that on $\phi$ is:
\begin{equation}
\sigma^2_\phi=\frac{Q^2\sigma^2_U+U^2\sigma^2_Q}{4(Q^2+U^2)^2}+\sigma^2_{\phi\,{\rm CAL}} \,,
\end{equation}
where, again, the calibration error on the polarisation angle, $\sigma_{\phi,{\rm CAL}}$, is added in quadrature. Under the hypothesis that the calibration error affects equally the $Q$ and $U$ parameters ($\sigma_{Q, {\rm CAL}}\simeq\sigma_{U, {\rm CAL}} = \sigma_{\rm CAL}$),
$\sigma_{P,{\rm CAL}}\simeq 0.1 P$ gives $\sigma_{\rm CAL}= 0.1 P/\sqrt{2}$.
Finally:
\begin{equation}
\sigma^2_{\phi,{\rm CAL}}\simeq \frac{1}{4(Q^2+U^2)}2\sigma^2_{\rm CAL}\simeq (0.05\,\mbox{rad})^2 \,,
\end{equation}
%
implying $\sigma_{\phi\,{\rm CAL}}\simeq 3\deg$. This estimation turns out to be consistent with differences in the polarisation angle obtained by using different calibrators.

\subsection{Imaging and flux density measurements}

To check our assumption of target compactness at all frequencies and the flux density extraction we created maps for each Stokes parameter by means of the standard MIRIAD procedure. We adopted the natural weighting, the standard for point sources, to ensure the lowest noise level. We used multi-frequency synthesis imaging and a standard H\"ogbom algorithm.


\noindent The detection threshold was set at $5\sigma$ in both $I$ and $P$. The $\sigma_{I_{\rm map}}$ was derived from the $I$ image region with no emission.

We use the task UVFLUX for flux density measurements (except for a few cases described below). This task is designed for point-like sources and works directly on visibilities, minimizing artifacts and phase-instabilities which may affect more image-based measurements. Since our sample excludes extended sources (at least up to $20\,$GHz), we expect that flux densities provided by UVFLUX and by IMSTAT (the MIRIAD task for flux density estimation from imaging) show differences no greater than $\simeq 10\%$ in total intensity (due to residual phase instabilities in the images). Moreover, we assumed a source pointing accurate enough to keep objects in the phase center of the $uv$-plane. The latter hypothesis is supported by the fact that the PACO catalogue is drawn from the AT20G one, for which all positions are known with an accuracy $\lesssim 1\,$arcsec (Murphy et al. 2010).

Afterwards, we compared the IMSTAT image peak (in mJy/beam) and the UVFLUX flux density estimation (for all the objects for which the uv-coverage allows imaging, i.e. $\simeq 90\%$ of the cases) to reveal, in both I and P, whether an extended or displaced (from the phase centre) component is present. The median discrepancies, $\Delta I/I$ and $\Delta P/P$, are $\simeq 2.7\%$ and $\simeq 4.3\%$, respectively. In total intensity the relative discrepancy is $> 10\%$ for $\simeq 4\%$ of the images. These large discrepancies are registered for frequencies $> 24\,$GHz; they reach maximum values between $20\%$ and $25\%$ for $2$ objects (AT20GJ080633-711217 and AT20GJ080649-610131, respectively) at $38\,$GHz. In all the cases the image peak values are higher than the UVFLUX estimation. A visual inspection of the images confirms that the sources are barely resolved.

\begin{figure}
\includegraphics[width=0.49\columnwidth]{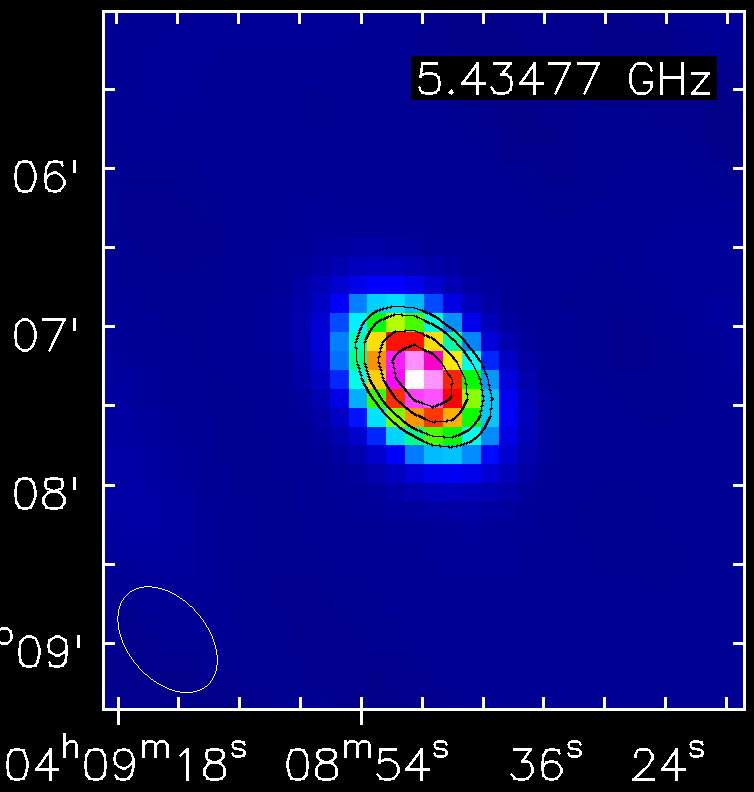}
\hspace{-0.1cm}
\includegraphics[width=0.50\columnwidth,height=0.51\columnwidth]{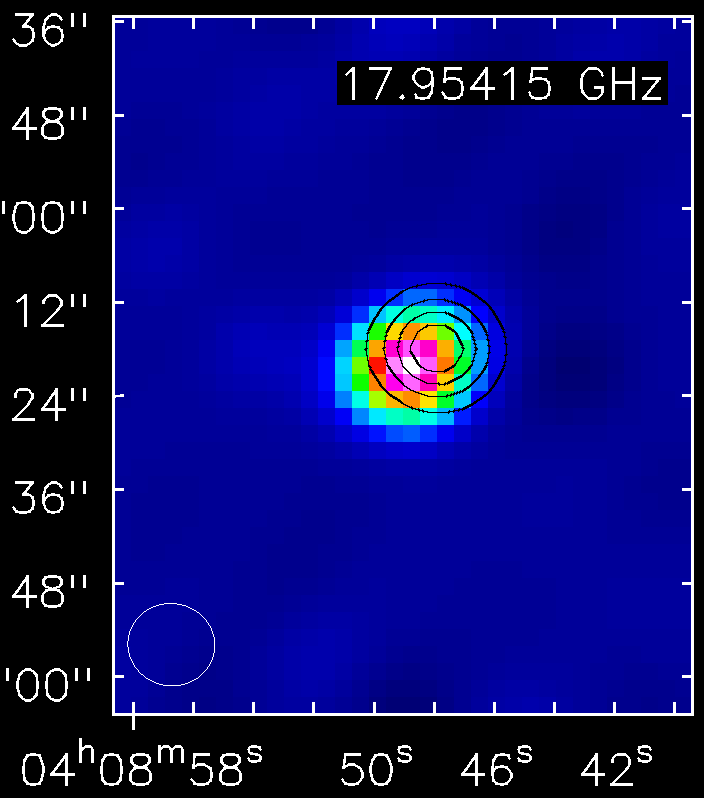}
\hspace{-0.1cm}
\includegraphics[width=0.49\columnwidth]{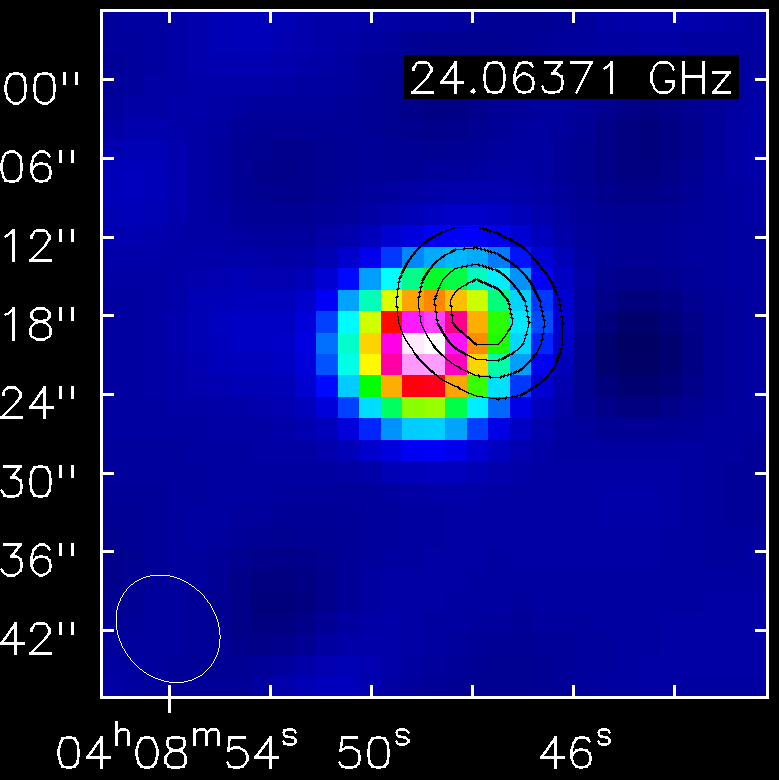}
\hspace{-0.1cm}
\includegraphics[width=0.50\columnwidth]{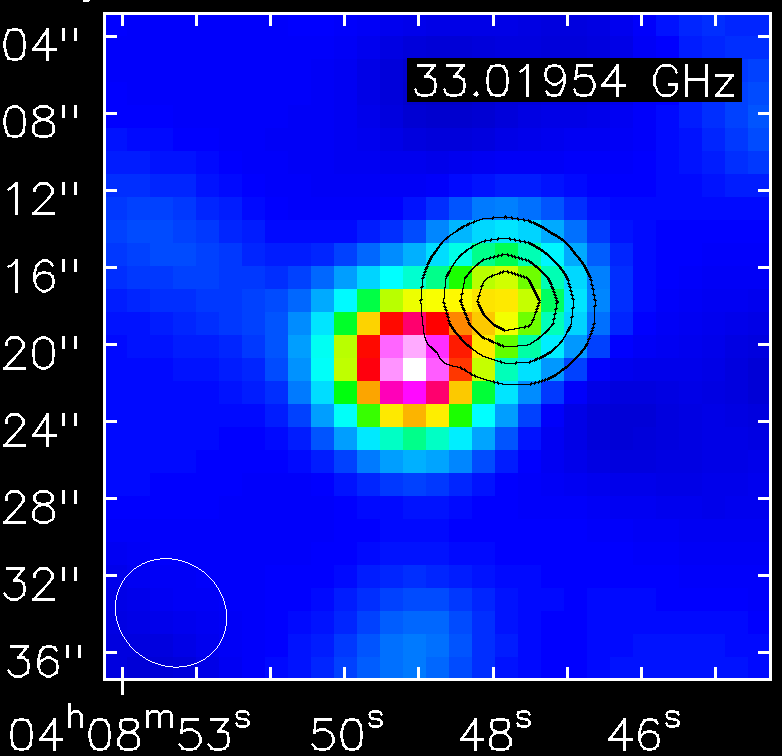}
\hspace{-0.1cm}
\includegraphics[width=0.49\columnwidth,height=0.51\columnwidth]{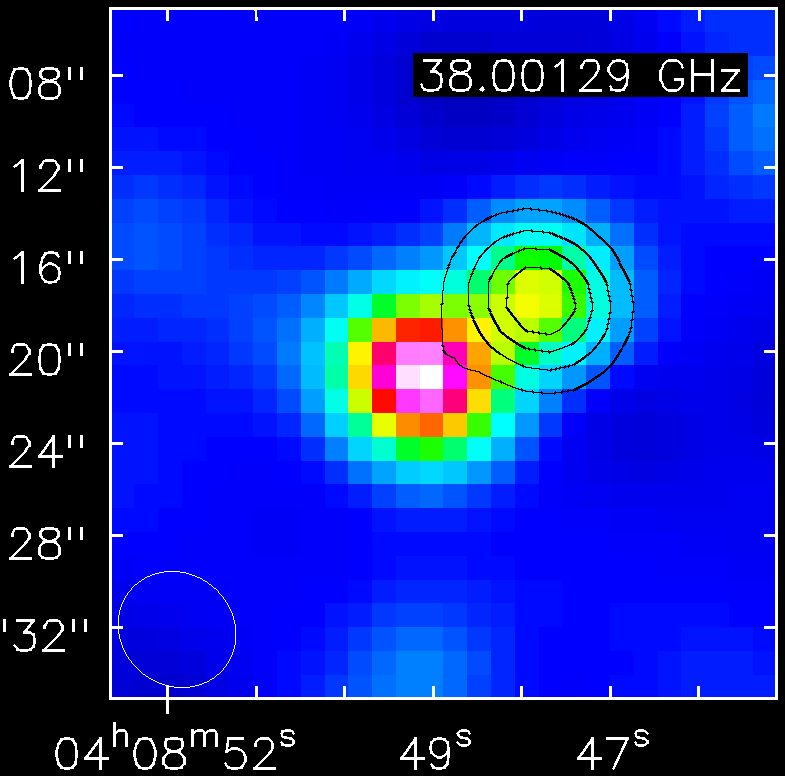}
\hspace{-0.1cm}
\includegraphics[width=0.50\columnwidth,height=0.51\columnwidth]{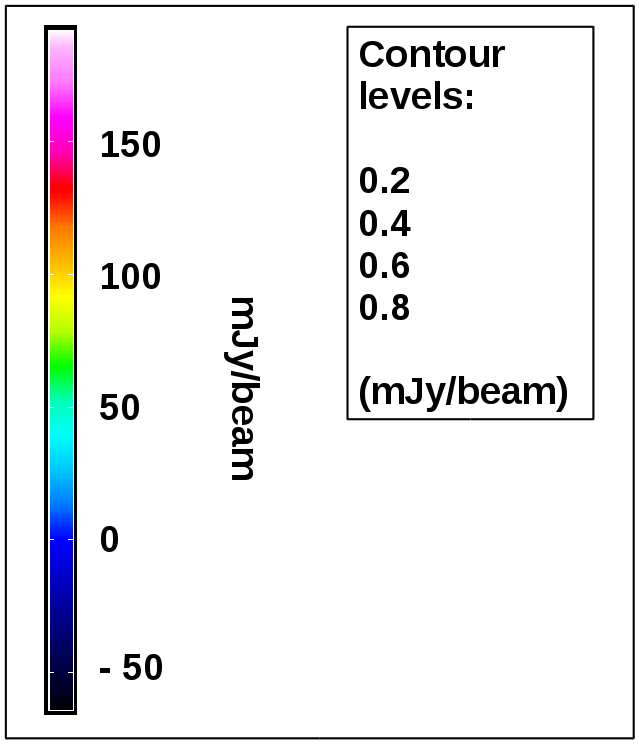}
\caption{Maps in total intensity (colours) and polarisation (contours) for AT20GJ0408-7507 at $5.5$, $18$, $24$, $33$ and $38\,$GHz. The two-lobe structure is resolved at the higher frequencies. The eastern lobe appears to be strongly depolarised.}
\label{fig:j0408}
\end{figure}

In polarisation the fractional discrepancy is $> 15\%$ in $\simeq 11\%$ of the available images and $>20\%$ in $\simeq 8\%$ of the cases. In particular, there is one object, AT20GJ040848-750720, for which the excess of flux density (with respect to UVFLUX) measured in correspondence of the peak in the image reaches $\simeq 361\%$ at $33\,$GHz and $\simeq 352\%$ at $38\,$GHz. We show images for this source (see Fig.~\ref{fig:j0408}) at all the $5$ frequencies at which they could be obtained in total intensity. Polarised emissions are instead displayed by contour levels. It can be seen that the object is marginally resolved at the lower frequencies. At the higher frequencies a second component appears and the polarised emission mainly comes from it. According to Morganti et al. (1999) AT20GJ0408-7507 is a bright FRII radio galaxy at $z\simeq 0.7$, dominated by two bright lobes. Both lobes have high depolarization, slightly higher in the eastern one.

Given the good matching between flux densities from imaging and from visibilities for point-like objects (both in total intensity and in polarisation), we decided to integrate over a suitable region on $I$ images to recover the total intensity flux densities in case of slightly resolved objects. In case of the flux discrepancies in polarisation, since images reveal an emerging point-like component displaced from the phase center, we estimated the flux density by considering the peak of the $P$ image.

The complete catalogue (flux densities and fitting parameters, polarisation fractions and angles) for the observations discussed in the present paper is available online, as supplementary material.

\section{Data analysis}
\label{dataanaly}

\begin{figure}
	\includegraphics[width=\columnwidth]{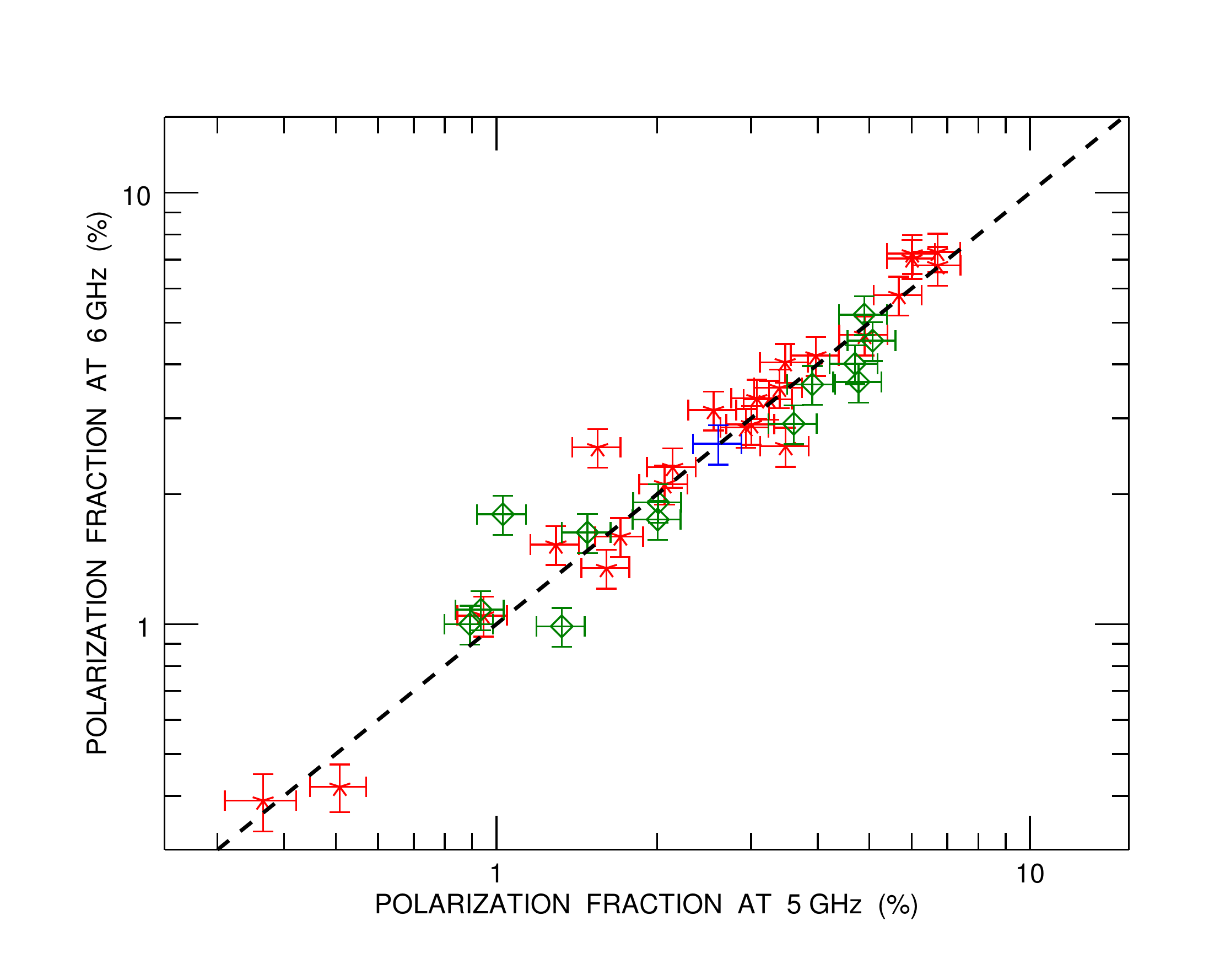}
	\includegraphics[width=\columnwidth]{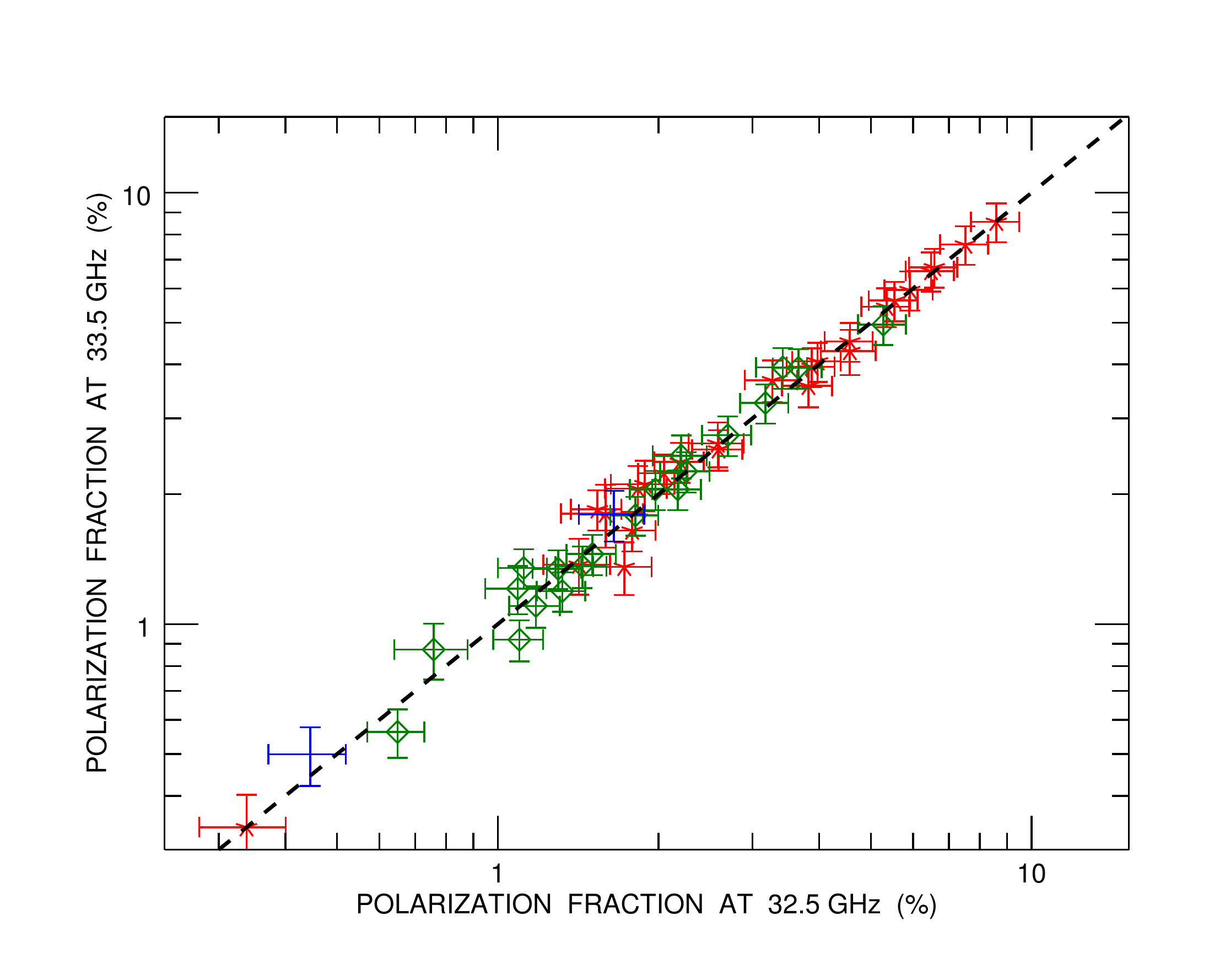}
    \caption{Comparison of the fractional polarisation measured in the two $1\,$GHz-wide sub-bands of the $2\,$GHz-wide bands at $5.5\,$GHz (upper panel) and at $33\,$GHz (lower panel). Red asterisks, blue $+$ signs and green diamonds refer to steep--spectrum, flat--spectrum and peaked--spectrum sources, respectively (the spectral classification is described in Sect.~\ref{sec:4.2}). The bisector is shown as a dashed line.}
    \label{fig:PPoldiag}
\end{figure}

We obtained $5\,\sigma$ detections in polarisation, with a median error of $\simeq 0.6\,$mJy, for $\simeq 91\%$ of our sources. The detection rate is nearly uniform across the observed frequencies ($48$ sources detected at $5.5\,$GHz and $49$ detected at $38\,$GHz). To check whether our data are affected by intra-band depolarisation at any frequency we have subdivided each $2\,$GHz-wide band into two $1\,$GHz-wide sub-bands and compared the polarisation degrees measured in the sub-bands. As illustrated by the examples shown in Fig.~\ref{fig:PPoldiag} no systematic difference was found.

\subsection{Fitting procedures}
\label{fit}

To properly fit source spectra we start considering a double power law represented by the expression
\begin{equation}
S(\nu)=\frac{S_0}{\left(\frac{\nu}{\nu_0}\right)^{-a}+\left(\frac{\nu}{\nu_0}\right)^{-b}},
\end{equation}
\noindent or by a concave version of it (needed in two cases)
\begin{equation}
S(\nu)=S_0\left(1-\frac{1}{\left(\frac{\nu}{\nu_0}\right)^{-a}+\left(\frac{\nu}{\nu_0}\right)^{-b}}\right),
\end{equation}

\noindent where $S_0$, $\nu_0$, $a$ and $b$ are free parameters. This function properly fits the total intensity data in $\simeq 96\%$ of the cases, confirming what found in previous works (e.g., Massardi et al. 2016b). In two cases (sources AT20GJ0546-6415 and AT20GJ0719-6218) the double power law provided a poor fit and we resorted to a triple power law model which requires 3 additional parameters, i.e. $S_1$, ${\nu_1}$ and $c$:
\begin{eqnarray}
S(\nu)&=&\frac{S_0}{\left(\frac{\nu}{\nu_0}\right)^{-a}+\left(\frac{\nu}{\nu_0}\right)^{-b}}+ \frac{S_1}{\left(\frac{\nu}{\nu_1}\right)^{-b}+\left(\frac{\nu}{\nu_1}\right)^{-c}}.
\label{eq:trplaw}
\end{eqnarray}
%
As for fitting procedures in polarisation, we need detections at no less than $6$ frequencies (over a maximum of $12$) in case of a double power law and at no less than $9$ frequencies for a triple power law. If a point source was not detected in one (or both) of the split frequency ranges of a band, we used the corresponding non-split detection, when available.  Given the small fraction ($\simeq 9\%$) of non-detections we did not consider the upper limits in doing the spectral fits. In only three cases (AT20GJ054641-641522, AT20GJ062524-602030 and AT20GJ075714-735308) we do not have detections at enough frequencies to get a proper fit.

About $90\%$ of our source spectra could be fitted with either a double power law or by a concave version of it (needed in $6$ cases).
The median reduced $\chi^2$ are $0.31$ and $0.82$ for I and P, respectively. This seems to confirm that our approach in error estimates is indeed conservative. Figure~\ref{fig:Spettri1} shows the data, the fitting curves and, when available, the previous PACO best epoch (2009-2010) observations in total intensity and the AT20G best epoch (2004-2008) observations in total intensity and in polarisation.

\begin{figure*}
\vspace{-1.5cm}
\centering
\includegraphics[scale=0.80]{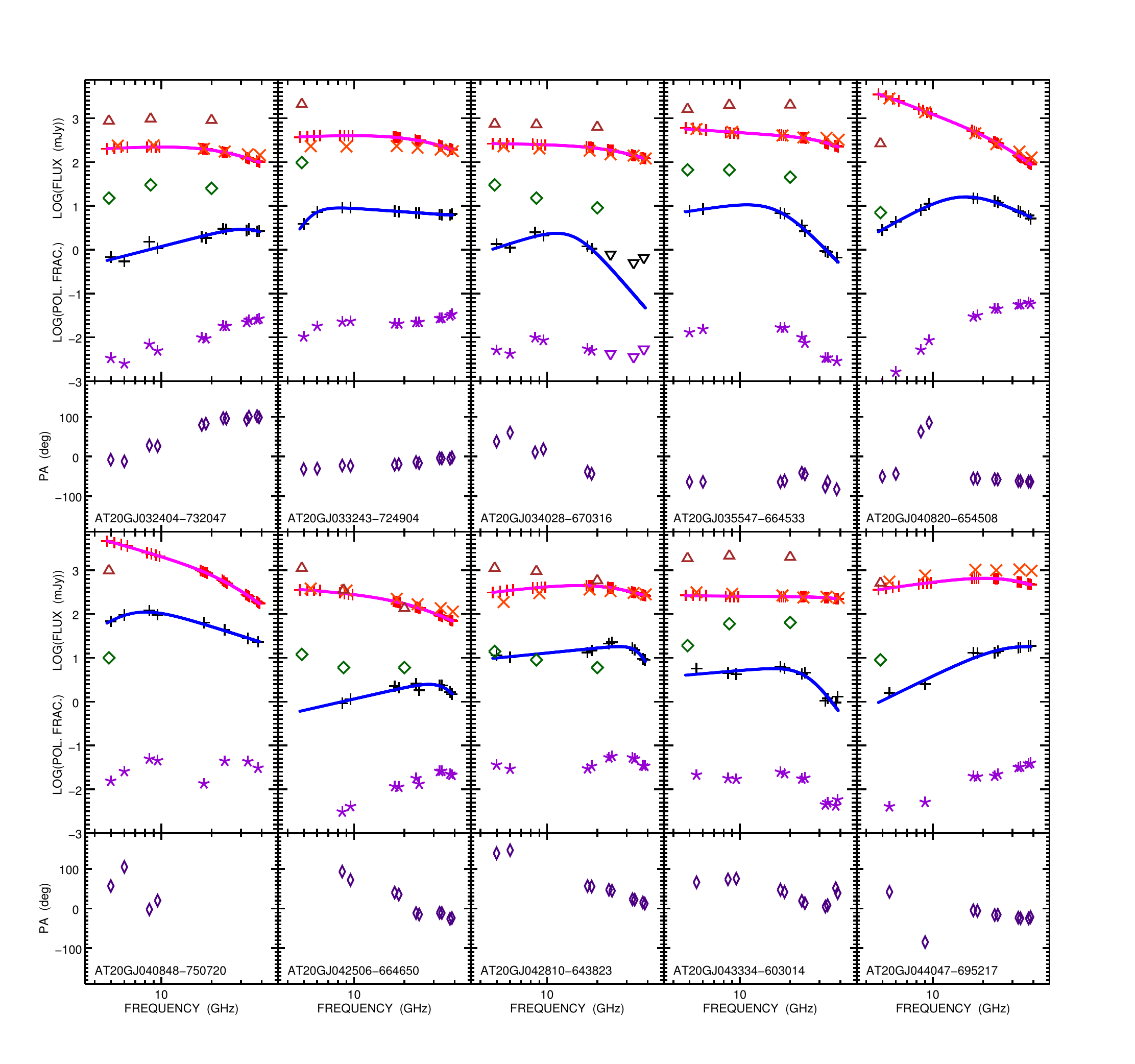}
\caption{Spectra in total intensity and polarisation, polarisation fraction and angle for the $53$ objects of the faint PACO sample, observed in September 2014. The error bars are not displayed since they are smaller than the symbols. {\bf Total intensity:} red pluses indicate ATCA September 2014 observations (each point represents a $512\,$MHz-wide sub-band) and the solid magenta lines show the fitting curves. The orange crosses show the median PACO flux densities (July 2009-August 2010) while the brown triangles show the AT20G observations (best epoch in 2004-2008). {\bf Polarisation (flux density):} black pluses refer to September 2014 observations and correspond to $1\,$GHz sub-bands or to the full $2\,$GHz band. Upper limits are shown as black downwards triangles. The solid blue lines show the fitting curves. The AT20G observations (best epoch in 2004-2008) are represented by green diamonds. {\bf Polarisation fraction:} violet asterisks refer to September 2014 observations. Upper limits are shown as downwards violet triangles. {\bf Polarisation angle:} this is displayed below each panel with flux densities and polarisation fraction. Indigo diamonds refer to the September 2014 campaign.}
\label{fig:Spettri1}
\end{figure*}
\addtocounter{figure}{-1}
\begin{figure*}
\vspace{-2.5cm}
\centering
\includegraphics[scale=0.80]{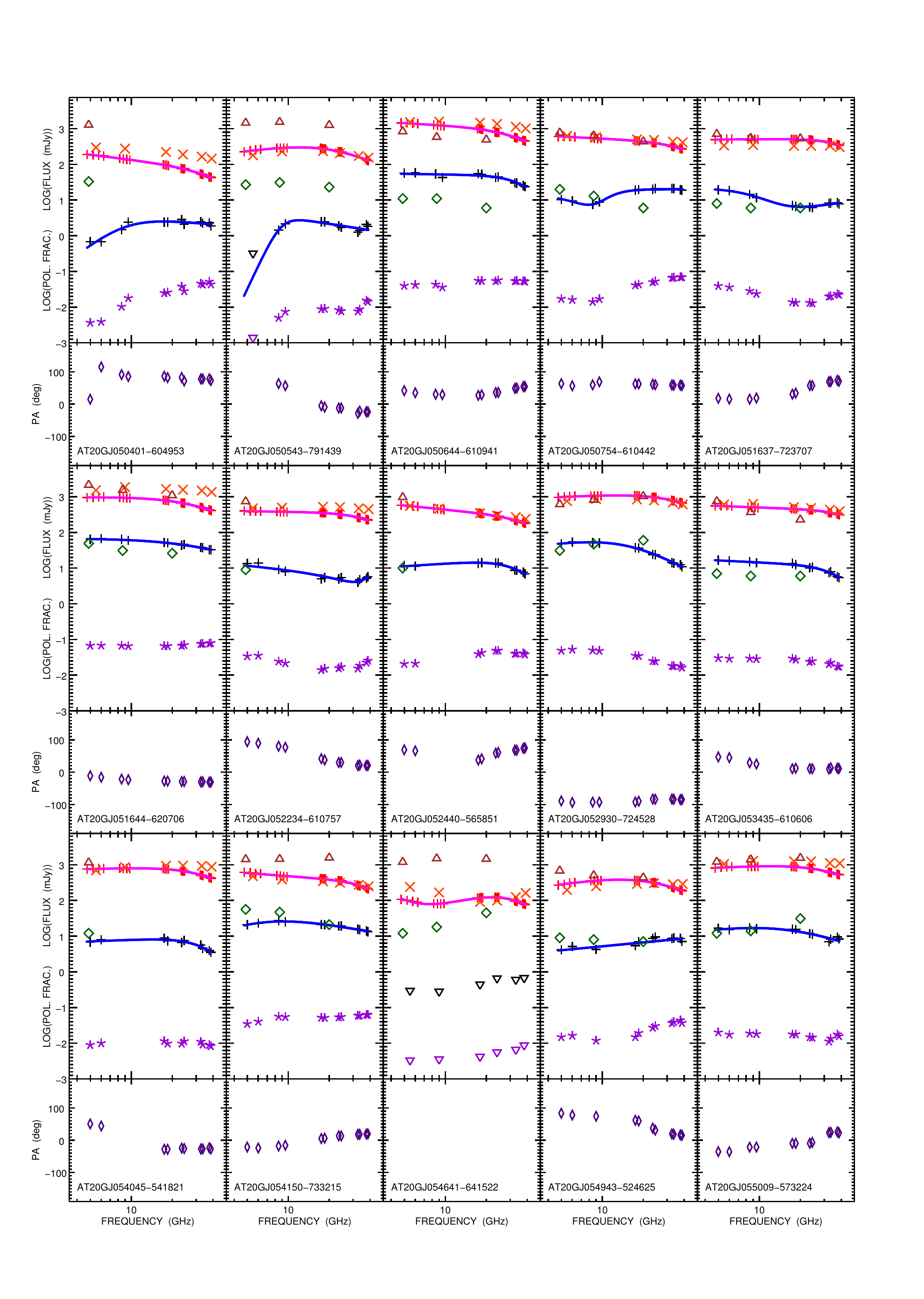}
\caption{Continued.}
\label{fig:Spettri2}
\end{figure*}
\addtocounter{figure}{-1}
\begin{figure*}
\vspace{-2.5cm}
\centering
\includegraphics[scale=0.80]{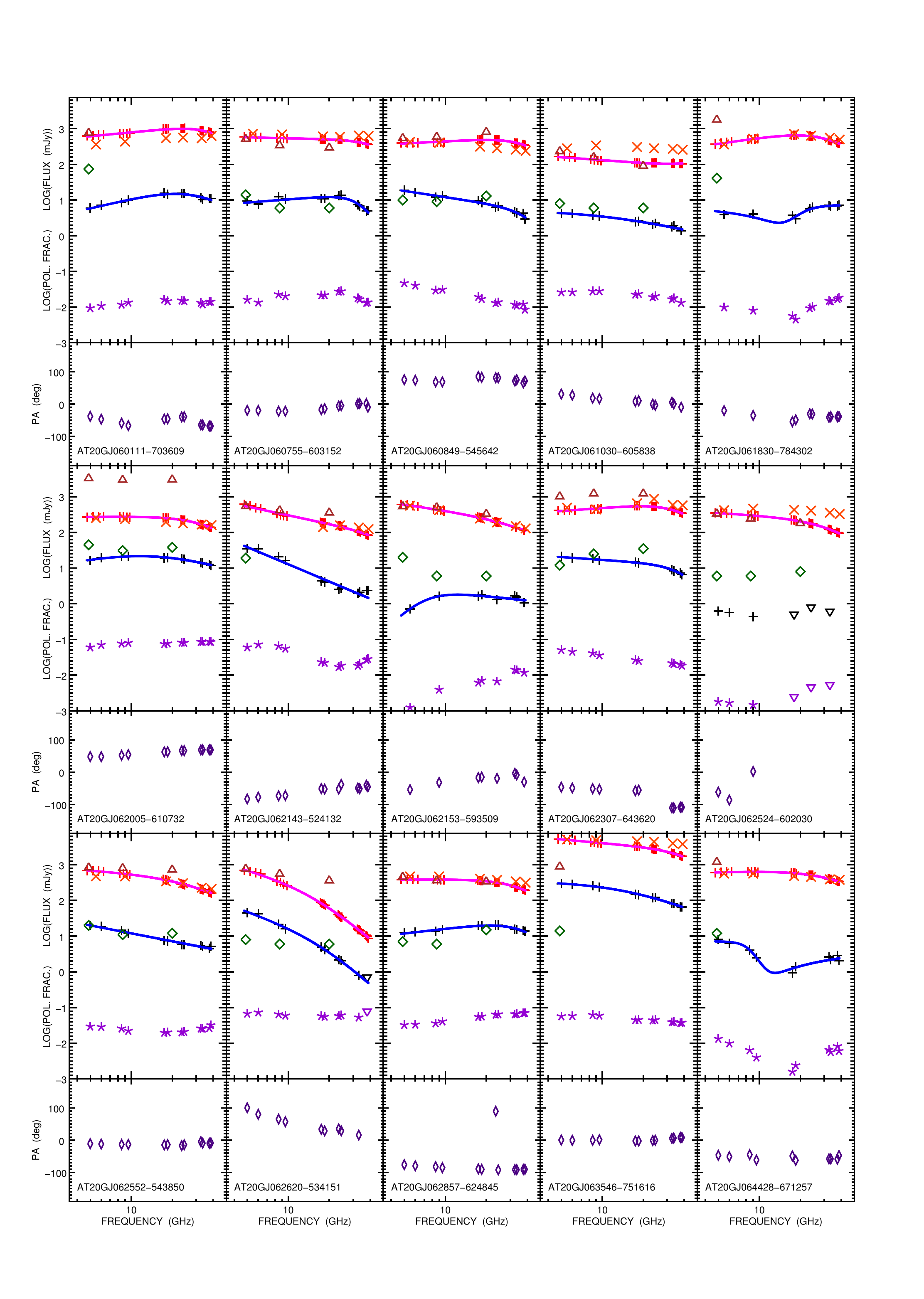}
\caption{Continued.}
\label{fig:Spettri3}
\end{figure*}
\addtocounter{figure}{-1}
\begin{figure*}
\vspace{-2.5cm}
\centering
\includegraphics[scale=0.80]{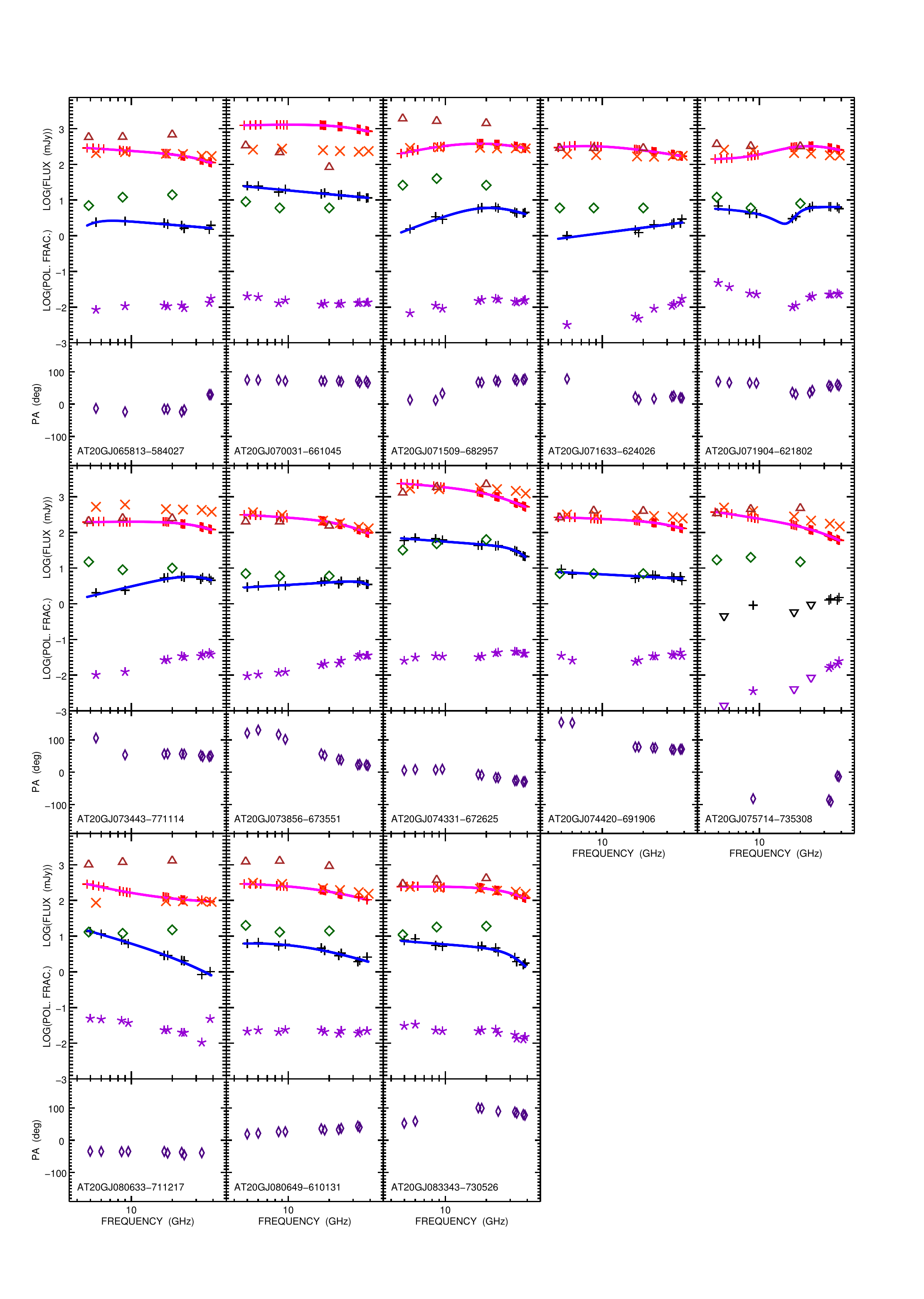}
\caption{Continued.}
\label{fig:Spettri4}
\end{figure*}

\begin{table}
\caption{Distribution of sources per spectral type in total intensity and in polarisation. The row ``NA'' refers to the three objects classified in total intensity but missing a spectral fit in polarisation. The last row reports the total for a given spectral class in total intensity, while the last column does the same in polarisation.}
\label{tab:Matnumsou}
\vspace{0.3cm}
\begin{tabular}{lcccccc}
\hline
Tot. Int. $\rightarrow$&(I)&(P)&(F)&(S)&(U)&\\
Pol. Int. $\downarrow$ & & & & & &\\
\hline
(I) & 0 & 1 & 0 & 0 & 0 & 1\\
(P) & 0 & 10 & 1& 13& 0 & 24\\
(F) & 0 & 3 & 0 & 2 & 0 & 5\\
(S) & 0 & 2 & 1 & 11& 0 & 14\\
(U) & 0 & 4 & 0 & 2 & 0 & 6\\
(NA)& 0 & 1 & 0&  2 & 0 & 3\\
\hline
\hline
&0&21&2&30&0&\\
\hline
\end{tabular}
\end{table}

\subsection{Spectral properties of the sample}
\label{sec:4.2}
We define the spectral index $\alpha_{\nu_1}^{\nu_2}$ between the frequencies $\nu_1$ and $\nu_2$ as:
\begin{equation}
\alpha_{\nu_1}^{\nu_2}=\frac{\log{\left (S(\nu_2)/S(\nu_1)\right )}}{\log{\left (\nu_2/\nu_1\right )}} \,,
\label{equ:SpeInd}
\end{equation}
where $S(\nu_1)$ and $S(\nu_2)$ are the flux densities at the two frequencies. To minimize fluctuations due to noise, spectral indices were calculated using the spectral fits in total intensity and in polarisation. We selected 5 frequencies (almost) equally spaced in a logarithmic scale, namely $5.5$, $10$, $18$, $28$ and $38\,$GHz.

\begin{table*}
\centering
\caption{First, second (median) and third quartiles of spectral indices in total intensity and in polarisation for different frequency ranges. We give values for the full sample and for the two main spectral classes, as classified in total intensity.}
\label{tab:Medvalspeind2}
\vspace{0.3cm}
\begin{tabular}{lcccc}
\hline
Tot. Int. & $5.5-10$ & $10-18$ & $18-28$ & $28-38\,$GHz\\
\hline
\begin{tabular}{l}Quart.\\\hline All (50)\\Steep (30)\\Peaked (21)\end{tabular}&\begin{tabular}{lcr}{\scriptsize 1}&2&{\scriptsize 3}\\\hline {\scriptsize -0.29}&$ -0.09$&{\scriptsize 0.07}\\{\scriptsize -0.44}&$-0.27$&{\scriptsize -0.15}\\{\scriptsize 0.05}&$0.12$&{\scriptsize 0.37}\\\end{tabular}&\begin{tabular}{lcr}{\scriptsize 1}&2&{\scriptsize 3}\\\hline {\scriptsize -0.44}&$-0.19$&{\scriptsize -0.06}\\{\scriptsize -0.60}&$-0.39$&{\scriptsize -0.25}\\{\scriptsize -0.10}&$-0.02$&{\scriptsize 0.25}\\\end{tabular}&\begin{tabular}{lcr}{\scriptsize 1}&2&{\scriptsize 3}\\\hline {\scriptsize -0.77}&$-0.53$&{\scriptsize -0.35}\\{\scriptsize -0.93}&$-0.66$&{\scriptsize -0.53}\\{\scriptsize -0.52}&$-0.39$&{\scriptsize -0.23}\\\end{tabular}&\begin{tabular}{lcr}{\scriptsize 1}&2&{\scriptsize 3}\\\hline {\scriptsize -1.22}&$-1.07$&{\scriptsize -0.88}\\{\scriptsize -1.31}&$-1.19$&{\scriptsize -1.03}\\{\scriptsize -1.10}&$-0.95$&{\scriptsize -0.80}\end{tabular}\\
\hline
\end{tabular}
\begin{tabular}{lcccc}
\hline
Pol. Int. & $5.5-10$ & $10-18$ & $18-28$ & $28-38\,$GHz\\
\hline
\begin{tabular}{l}Quart.\\\hline All (50)\\Steep (28)\\Peaked (20)\end{tabular}&\begin{tabular}{lcr}{\scriptsize 1}&2&{\scriptsize 3}\\\hline {\scriptsize -0.32}&$ 0.12$&{\scriptsize 0.48}\\{\scriptsize -0.29}&$0.03$&{\scriptsize 0.41}\\{\scriptsize -0.45}&$0.25$&{\scriptsize 0.87}\\\end{tabular}&\begin{tabular}{lcr}{\scriptsize 1}&2&{\scriptsize 3}\\\hline {\scriptsize -0.60}&$-0.26$&{\scriptsize 0.29}\\{\scriptsize -0.72}&$-0.29$&{\scriptsize 0.17}\\{\scriptsize -0.35}&$-0.03$&{\scriptsize 0.53}\\\end{tabular}&\begin{tabular}{lcr}{\scriptsize 1}&2&{\scriptsize 3}\\\hline {\scriptsize -0.81}&$-0.50$&{\scriptsize 0.00}\\{\scriptsize -1.07}&$-0.55$&{\scriptsize -0.34}\\{\scriptsize -0.58}&$0.01$&{\scriptsize 0.62}\\\end{tabular}&\begin{tabular}{lcr}{\scriptsize 1}&2&{\scriptsize 3}\\\hline {\scriptsize -1.63}&$-0.94$&{\scriptsize -0.32}\\{\scriptsize -1.74}&$-1.21$&{\scriptsize -0.61}\\{\scriptsize -1.24}&$-0.47$&{\scriptsize 0.29}\end{tabular}\\
\hline
\end{tabular}
\end{table*}

We classified the sources according to their $\alpha_{5.5}^{10}$ and $\alpha_{28}^{38}$ spectral indices. The choice of these frequency intervals follows from the fact that the majority of spectral peaks occur around $10-20\,$GHz.

We defined as flat-spectrum (F) sources those with $-0.5<\alpha_{5.5}^{10}<0.5$ and $-0.5<\alpha_{28}^{38}<0.5$. Sources outside these spectral index ranges were sub-divided as:
\begin{itemize}
    \item steep-spectrum (S), if $\alpha_{5.5}^{10}<0$ and $\alpha_{28}^{38}<0$;
    \item inverted-spectrum (I), if $\alpha_{5.5}^{10}>0$ and $\alpha_{28}^{38}>0$;
    \item peaked-spectrum (P), if $\alpha_{5.5}^{10}>0$ and $\alpha_{28}^{38}<0$;
   \item upturning-spectrum (U), if $\alpha_{5.5}^{10}<0$ and $\alpha_{28}^{38}>0$.
\end{itemize}

\noindent Table~\ref{tab:Matnumsou} reports the distribution among these spectral types for the complete sample. In both cases, the most populated spectral types are P and S. However, there is only a weak correspondence between spectral types in total intensity and in polarisation: less than half P- or S-type sources in total intensity have the same classification in polarisation. On the other hand, the statistics is quite low so that any firm conclusion is premature.

In Table~\ref{tab:Medvalspeind2} we list the quartiles of spectral indices, both in total intensity and in polarisation, for the full sample and for the two most populated  spectral classes: steep- and peaked-spectrum sources. Also in the polarisation case we kept the spectral classification done in total intensity.

The median spectral indices become steeper and steeper with increasing frequency. In total intensity they become, at $\nu \gtrsim 28\,$GHz, steeper than the typical spectral index of classical low-frequency steep-spectrum sources ($\alpha \simeq -0.75$), even though almost all these sources are classified as ``flat-spectrum'' at low frequencies. A similar trend is found in the polarized flux density, although the high frequency steepening is somewhat less pronounced.

A less steep median high-frequency spectral index in total intensity ($\alpha_{30}^{40}\simeq -0.69$) was reported  by Bonavera {\it et. al} (2011) for the full PACO ``faint''  sample of 159 sources. We have checked that the difference can be accounted for by the revision of the model for the reference flux density calibrator, PKS1934-638. Using the previous model, adopted by Bonavera et al. (2011) but less accurate for frequencies $>11\,$GHz, we get, for our sample, a median spectral index of $-0.68\pm 0.05$, consistent with the results by Bonavera et al. (2011).
Somewhat flatter spectral indices in polarisation, compared to total intensity, were also found by Massardi et al. (2013) for the ``bright'' AT20G sample ($S_{20\,\rm GHz}>500\,$mJy); for these objects, however, the data were limited to $\nu \le 18\,$GHz.

\begin{figure*}
\includegraphics[width=\columnwidth]{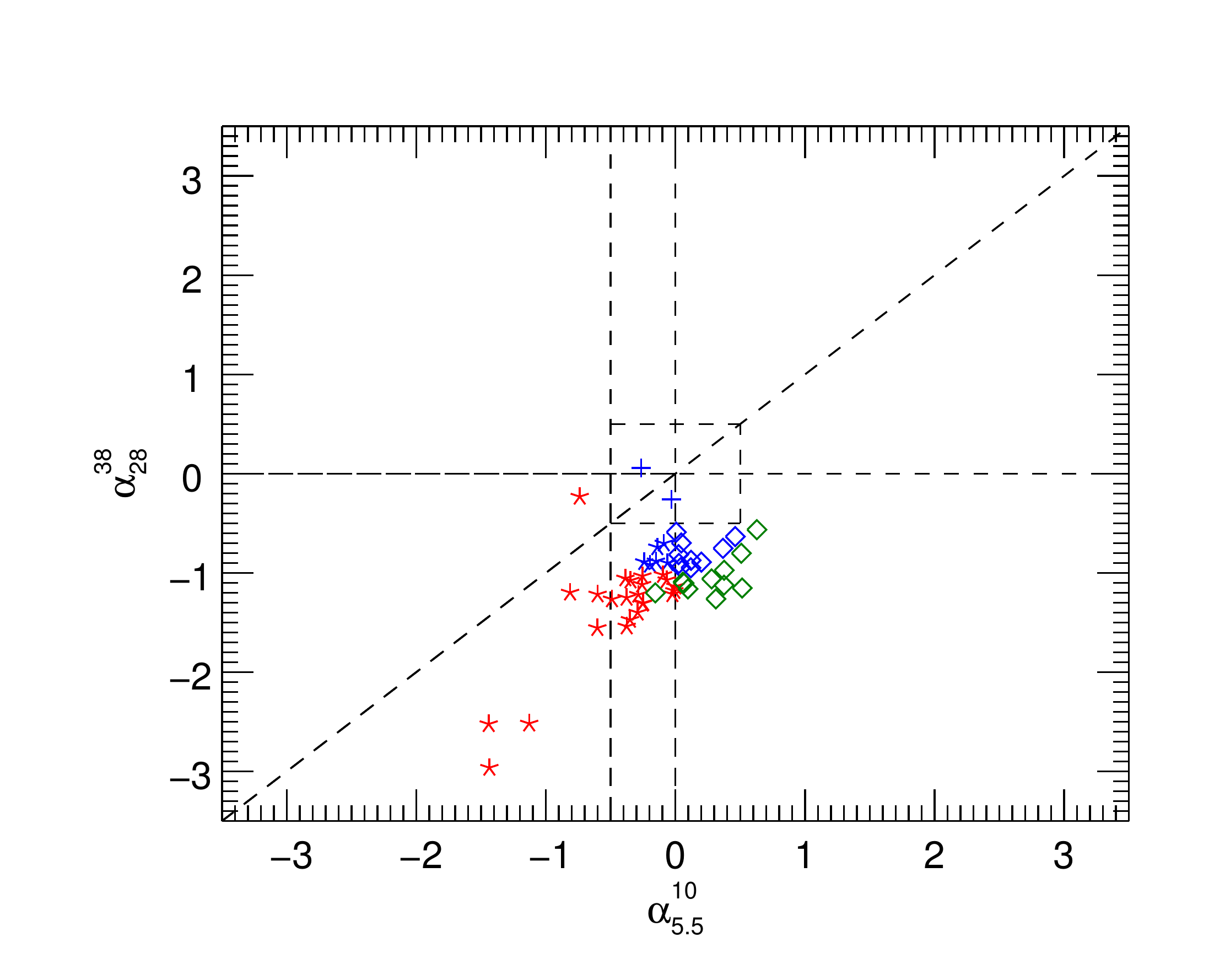}
\includegraphics[width=\columnwidth]{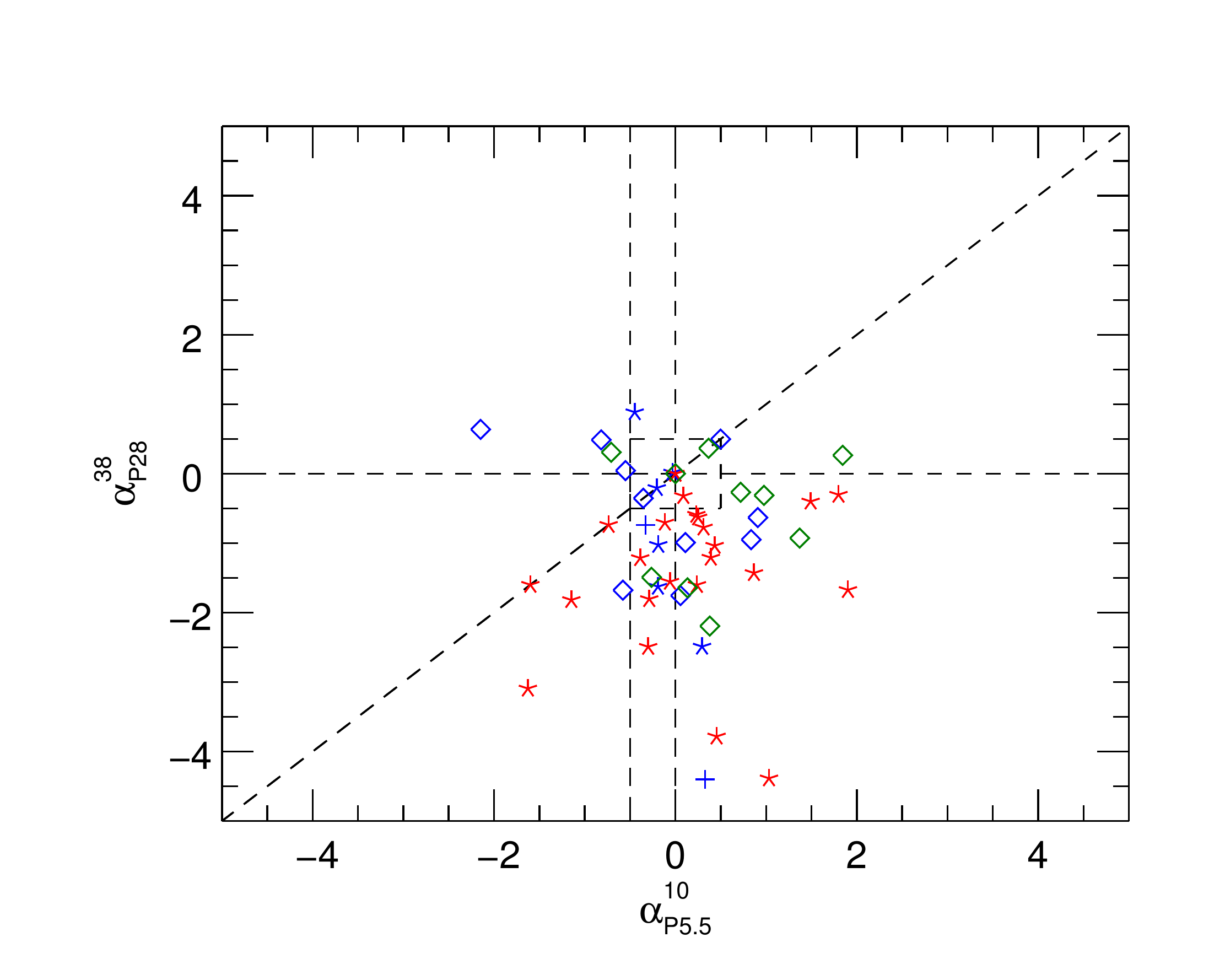}
\caption{Radio colour-colour diagrams for (from left to right) total intensity and polarised flux density. Symbols identify the spectral type in total intensity: pluses for flat-spectrum, asterisks for steep-spectrum, diamonds for peaked-spectrum. Colours refer to the spectral shape between  $5.5$ and $18\,$GHz: red for steep-spectrum, blue for flat-spectrum, and green for peaked-spectrum sources.}
\label{fig:CCLH}
\end{figure*}

\begin{figure*}
\includegraphics[width=\columnwidth]{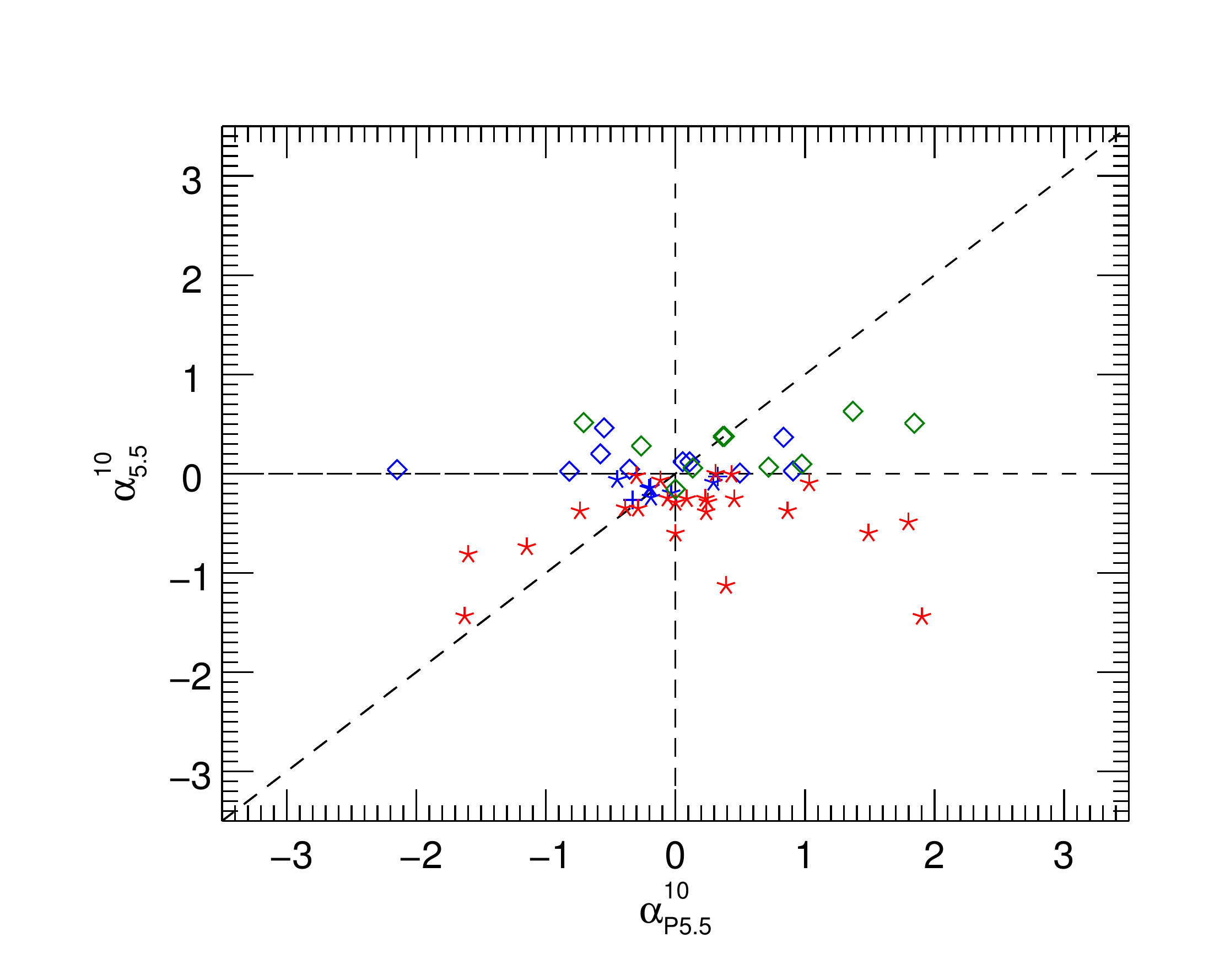}
\includegraphics[width=\columnwidth]{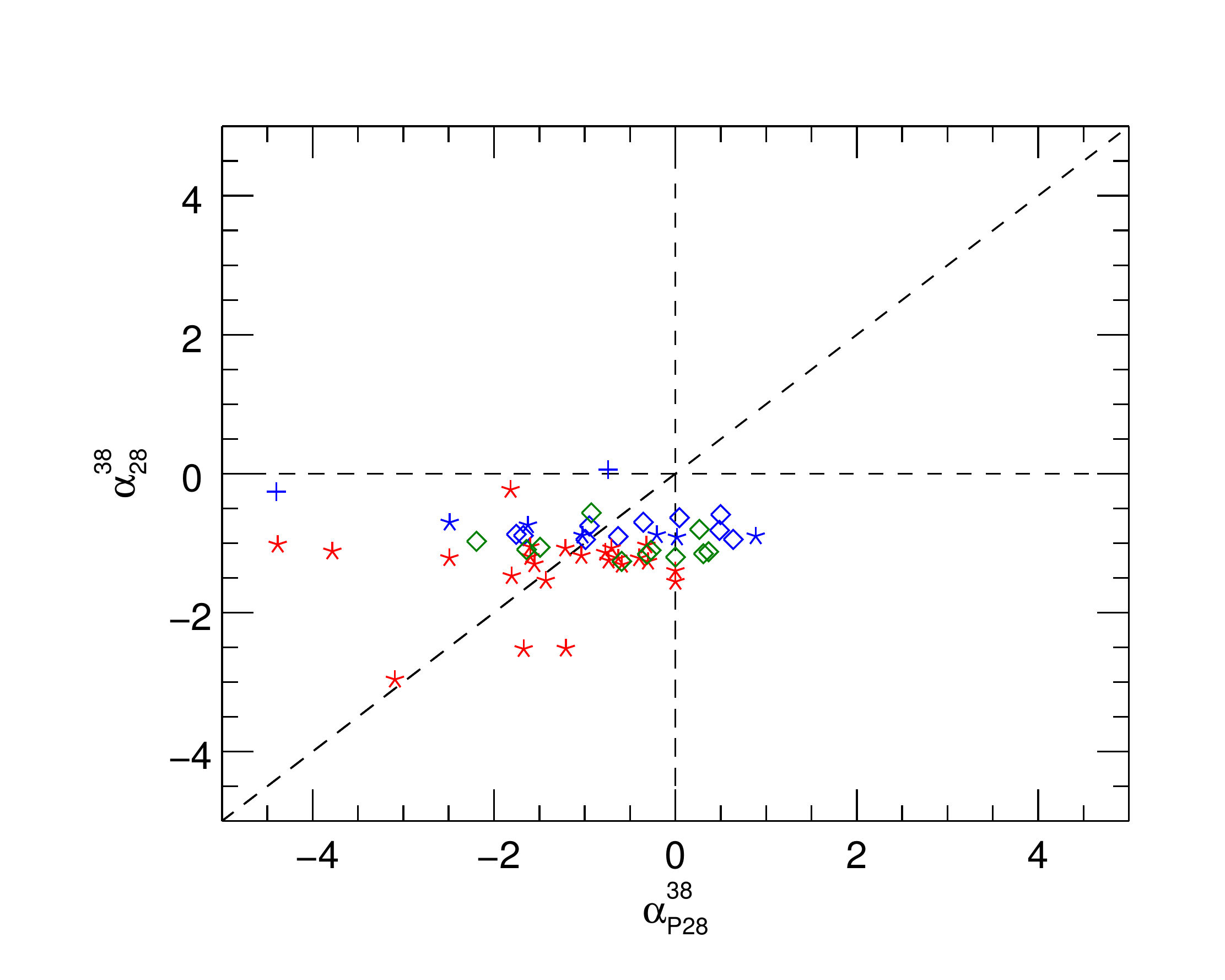}
\caption{Comparison between the spectral index in total intensity and in polarisation in the ranges $5.5-10\,$GHz (left) and $28-38\,$GHz (right). The meaning of symbols and colours is the same as in Fig.~\ref{fig:CCLH}.}
\label{fig:CCIP}
\end{figure*}

Figure~\ref{fig:CCLH} shows total intensity and polarisation colour-colour ($28-38\,$GHz {\it vs.} $5.5-10\,$GHz) plots. As explained in the caption, the symbols correspond to the spectral classification in total intensity mentioned above while the colours correspond to the spectral shape in the $5.5-18\,$GHz range. The left-hand panel shows a clear segregation of sources of each colour. The median spectral steepening $\Delta \alpha=\alpha_{28}^{38}-\alpha_{5}^{10}$ varies from $-0.86$, the value obtained both for the red (steep-spectrum) and blue (flat-spectrum), to $-1.41$ for the green (peaked-spectrum) ones. No analogous segregation is seen when considering the spectral indices of the polarised emission (right-hand panel).
Figure \ref{fig:CCIP} shows that there is no significant correlation between the $5.5-10\,$GHz (left-hand panel) or the $28-38\,$GHz (right-hand panel) spectral indices in total intensity and in polarisation. This implies that the polarisation degree of individual sources varies with frequency, as pointed out by previous analyses. Here we reiterate this result with higher accuracy and extend it to higher frequencies and fainter flux densities. On the other hand, as discussed in the following sub-section, we do not find any significant frequency dependence of the mean polarisation degree.

\subsection{Polarisation fraction}

Polarisation data, provided by the NVSS survey at $1.4\,$GHz, brought Tucci et al. (2004) to argue that Faraday depolarisation can be significant up to $\simeq 10\,$GHz. Furthermore, especially for compact objects, magnetic fields may be expected to be increasingly ordered for regions closer and closer to the central AGN, that are likely to dominate at the highest radio frequencies.

Both elements would lead to an increase of the polarisation degree with rest-frame frequency. According to Tucci et al. (2004), the mild anti-correlation between polarisation degree and 1.4 GHz flux density, detected for steep-spectrum sources in the NRAO VLA Sky Survey (NVSS), could be due to an increase of the mean redshift of sources (hence of the rest-frame frequency) with decreasing flux density. However no trend of the polarisation degree with 1.4 GHz flux density was detected for NVSS flat-spectrum sources and the reality of the anti-correlation for steep-spectrum sources was questioned by Hales et al. (2014).

No significant variations of the polarisation degree with either flux density or with frequency were found by Massardi et al. (2013) and Battye et al. (2011). Higher polarization degrees at mm wavelengths were reported by Agudo et al. (2010, 2014) and by Sajina et al. (2011). However, the former studies an higher-frequency selected sample ($S_{86\rm GHz}> 0.9\,$Jy), while the latter focuses on a fainter population ($S_{20 \rm GHz} > 40\,$mJy) and their high frequency results are likely biased by not having taken into account non detections (Tucci \& Toffolatti 2012).

\begin{figure}
\includegraphics[width=\columnwidth]{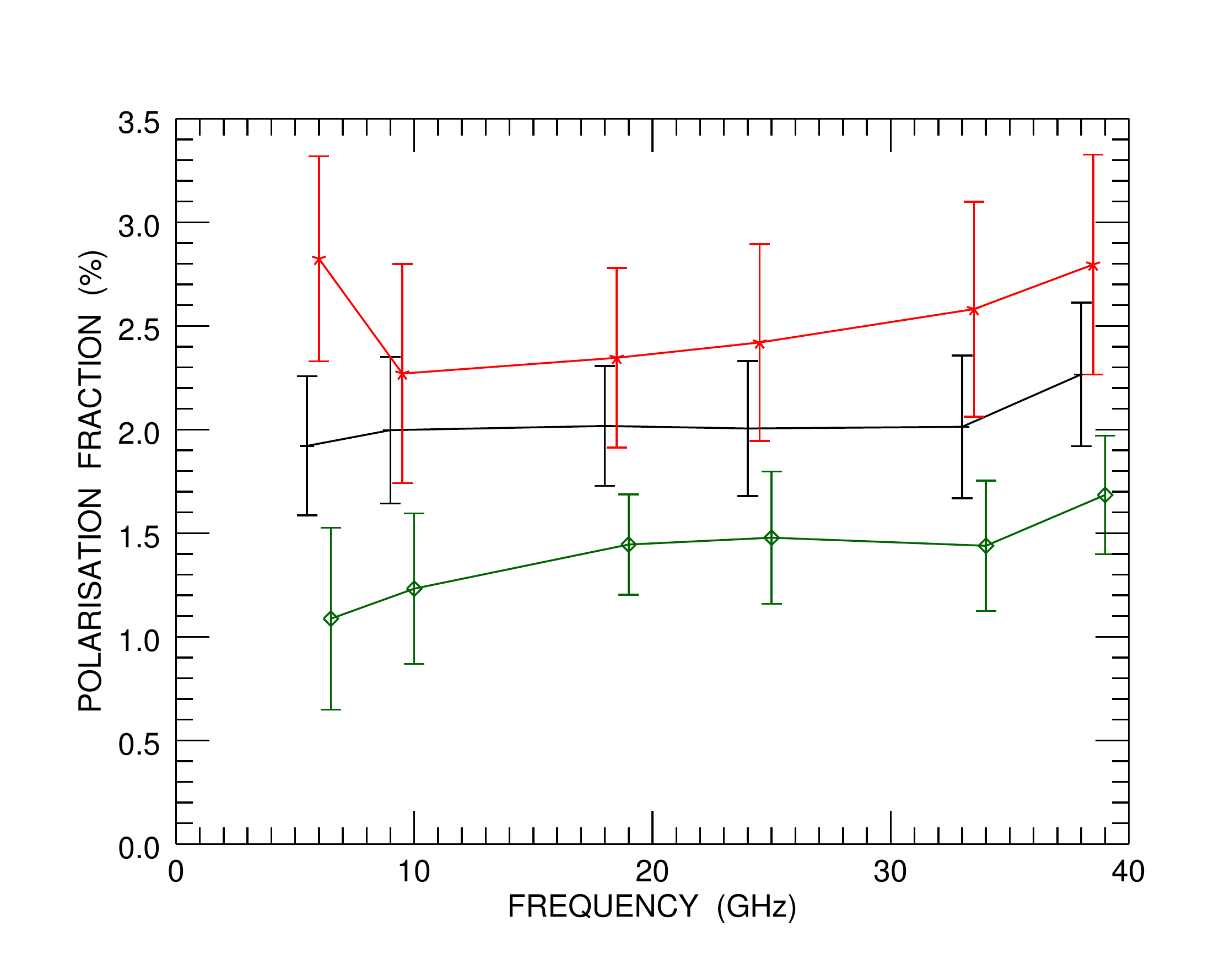}
\caption{Median polarisation fraction behaviour with frequency (at $5.5$, $9$, $18$, $24$, $33$ and $38\,$GHz) for all the sources (black), for steep sources (red) and for peaked ones (green). The errors on median values are given by $1.253\, {\rm rms}/\sqrt{N}$, where rms is the standard deviation around the mean and $N$ is the number of the data (at a given frequency) for a given class of objects (cf.~Arkin \& Colton 1970).}
\label{fig:PPolVsfreq}
\end{figure}

We addressed this issue with a slightly better combination of flux density and frequency ranges than the previous studies of samples selected at high frequency. The high detection rate in polarisation of our complete sample and the use of survival analysis techniques to take into account upper limits ensure that the selection bias must be negligible. The significance of the rejection of the null hypothesis (no correlation of the polarisation degree with frequency) was tested computing the distributions of polarisation degrees at each frequency by means of the Kaplan-Meier estimator included in the ASURV package (Lavalley, Isobe \& Feigelson 1992). Then the distributions ({\it cf.}~Tab.~\ref{tab:PPolVsfreq}) were compared using the Kolmogorov-Smirnov test included in the same package. The distributions at $5.5$ and $38\,$GHz are consistent with being drawn from the same parent distribution (see also Fig.~\ref{fig:PPolVsfreq}) in all the cases (all the sample, steep sources only and peaked sources only) with a rejection level $< 1 \sigma$. However, we are aware that an hypothetical trend may be easily washed out given the dispersions of polarisation fractions and the limited size of the sample. To justify this, we point out the variety of behaviours in polarised spectra ({\it cf.} Tab.~\ref{tab:Matnumsou}) which results in higher dispersions shown by spectral indices in polarisation with respect to total intensity ones ({\it cf.} Tab.~\ref{tab:Medvalspeind2}), hence in higher scatters of polarisation degrees. Again, we warn the reader that the flux density range spanned by our sample is quite limited.


As shown by the Table~\ref{tab:PPolVsfreq} and by Fig.~\ref{fig:PPolVsfreq} the median polarisation fraction is generally slightly higher for steep-spectrum  sources than for the peaked-spectrum ones. The last line of Table~\ref{tab:PPolVsfreq} shows, for each frequency, the probability that the two source types are drawn from the same parent population (null hypothesis), obtained from the Kolmogorov-Smirnov test. At 5.5 and 9.0 GHz this hypothesis can be rejected at a level between $2\,\sigma$ and $3\,\sigma$, while at higher frequencies the rejection level is between $4\,\sigma$ and $5\,\sigma$. Combining the data at all frequencies, the probability of the null hypothesis is $\sim 5\cdot 10^{-8}$ so that the hypothesis can be rejected at a $> 5\,\sigma$ significance.

\begin{table*}
\caption{First, second (median) and third quartiles of the polarisation fraction at each observed frequency given by the Kaplan-Meier estimator, taking into account the upper limits, for the full sample and for the steep- and peaked-spectrum sources. The last row reports probabilities for the null hypothesis (i.e. the two samples are drawn from the same parent distribution) given by the Kolmogorov-Smirnov test performed on the steep and peaked groups, considering together $5.5$ and $9\,$GHz, the $18-38\,$GHz frequency interval and all the frequencies, respectively.}
\label{tab:PPolVsfreq}
\begin{tabular}{cc}
\hline
Class.& frequencies (GHz)\\

\begin{tabular}{@{}c@{}}
\\
\hline
\\
All\\
Steep\\
Peaked\\
\hline
Prob.
\end{tabular}&\hspace{-0.3cm}
\begin{tabular}{@{}cccccc@{}}
5.5&\hspace{-0.3cm}9&\hspace{-0.3cm}18&\hspace{-0.3cm}24&\hspace{-0.3cm}33&\hspace{-0.3cm}38\\
\hline
\begin{tabular}{lcr}{\scriptsize 1}&2&{\scriptsize 3}\\\hline {\scriptsize 0.92}&$ 1.92$&{\scriptsize 3.52}\\ {\scriptsize 0.98}&$ 2.82$&{\scriptsize 3.74}\\{\scriptsize 0.47}&$ 1.09$&{\scriptsize 2.91}\end{tabular}&\hspace{-0.3cm}\begin{tabular}{lcr}{\scriptsize 1}&2&{\scriptsize 3}\\\hline {\scriptsize 0.89}&$ 2.00$&{\scriptsize 3.49}\\ {\scriptsize 1.03}&$ 2.27$&{\scriptsize 4.13}\\ {\scriptsize 0.61}&$ 1.24$&{\scriptsize 2.32}\end{tabular}&\hspace{-0.3cm}\begin{tabular}{lcr}{\scriptsize 1}&2&{\scriptsize 3}\\\hline {\scriptsize 1.10}&$ 2.02$&{\scriptsize 2.99}\\ {\scriptsize 1.55}&$ 2.35$&{\scriptsize 4.27}\\ {\scriptsize 0.90}&$ 1.45$&{\scriptsize 1.92}\end{tabular}&\hspace{-0.3cm}\begin{tabular}{lcr}{\scriptsize 1}&2&{\scriptsize 3}\\\hline {\scriptsize 1.28}&$ 2.01$&{\scriptsize 3.33}\\ {\scriptsize 1.59}&$ 2.42$&{\scriptsize 4.67}\\ {\scriptsize 1.04}&$ 1.48$&{\scriptsize 2.14}\end{tabular}&\hspace{-0.3cm}\begin{tabular}{lcr}{\scriptsize 1}&2&{\scriptsize 3}\\\hline {\scriptsize 1.26}&$ 2.01$&{\scriptsize 3.79}\\ {\scriptsize 1.69}&$ 2.58$&{\scriptsize 5.09}\\ {\scriptsize 1.14}&$ 1.44$&{\scriptsize 2.30}\end{tabular}&\hspace{-0.3cm}\begin{tabular}{lcr}{\scriptsize 1}&2&{\scriptsize 3}\\\hline {\scriptsize 1.37}&$ 2.27$&{\scriptsize 3.88}\\ {\scriptsize 1.53}&$ 2.80$&{\scriptsize 4.68}\\ {\scriptsize 1.34}&$ 1.69$&{\scriptsize 2.53}\end{tabular}\\
\hline
{\scriptsize ($5.5-9\,$GHz)}&\hspace{-1cm}{\scriptsize $0.0355238$}&{\scriptsize ($18-38\,$GHz) }&\hspace{-1cm}{\scriptsize $8.75641\cdot 10^{-7}$}& {\scriptsize (All freqs.)}&\hspace{-1cm}{\scriptsize $5.39317\cdot 10^{-8}$}
\end{tabular}\\
\hline
\end{tabular}
\end{table*}



\subsection{Rotation measures}
The polarisation angle was calibrated setting the parameter ``xycorr'' in the MIRIAD task ATLOD which applies phase corrections provided by a noise diode mounted on one antenna feed. Partridge et al. (2016) found that the polarisation angles measured by ATCA in this way agree with those measured by \textit{Planck} to within $\pm 2\,\deg$.

Our multi-frequency data have allowed us to perform a search for Faraday rotation. To this end we divided each $2\,$GHz band in $1\,$GHz sub-bands. We first checked the compatibility of the position angles measured for each sub-band with those measured for the full band. In almost all cases we found good agreement within the errors. Considering the sub-bands we have position angle measurements for each source at up to 12 wavelengths. In the presence of Faraday rotation the polarisation angle, $\phi$, has a simple dependence on the wavelength, $\lambda$, namely
\begin{equation}
\phi=\phi_0+RM \lambda^2,
\label{equ:PhiRotMea}
\end{equation}
where $\phi_0$ is the intrinsic polarisation angle and $RM$ is the rotation measure. We adopt the IDL\footnote{The ``Interactive Data Language'' (IDL) is a scientific programming language released by Harris Geospatial Solutions (http://www.harrisgeospatial.com).} procedure ``linfit'' for estimating the fitting parameters. We assume a fit is considered acceptable if the reduced $\chi^2 < 3$ and the associated probability level $p> 0.1$ (as suggested by the algorithm), which means that the computed parameters are believable. According to these criteria, our polarisation measurements show evidence of non-zero Faraday rotation only for $2$ sources considering the whole spectral range $5.5-38\,$GHz (AT20GJ051644-620706 and AT20GJ070031-661045). The derived values of the rotation measure are $(96\pm 13)\,{\rm rad}/{\rm m}^2$ and $(30\pm 13)\,{\rm rad}/{\rm m}^2$, respectively. The rest of the sample shows various and more complex behaviours: we notice a marked difference typically occurring between the lower frequencies ($5.5$ and $9\,$GHz) and the higher frequencies ($18$ and $24\,$GHz). This suggests different plasma conditions and/or magnetic field structures in the regions dominating the emission in the two frequency ranges. We defer a detailed analysis of this issue to a future paper, that will take advantage of a larger sample and of a wider spectral coverage obtained, in our most recent ATCA polarimetric observations.  

\section{Variability}
\label{variability}

Taking into account also the PACO and the AT20G measurements, the 53 sources analysed in this work have at least three epochs of observations in total intensity and two epochs in polarisation (no polarisation data from PACO). The AT20G data were taken between 2004 and 2008, the PACO data between July 2009 and August 2010 and our observations were done in September 2014. The typical global time span is of about eight years.

In combining the PACO measurements with the two other data sets we neglected the small difference in the central frequencies ($39$ vs $38\,$GHz) of the highest frequency channels. For each but three sources in our sample there is at least one PACO epoch for which all the six frequencies were observed. In several cases all the frequencies were observed two or three times. In many more cases we have repeated observations for only a subset of frequencies. Since the variability on timescales from a few months to $2-4$ years were discussed in previous works of our group (e.g. Massardi et al. 2016b), we focus here on longer timescales ($5$ to $8$ years).

We have considered only PACO observations of at least 4 frequencies, all carried out within several months. Multiple PACO observations of a sources at a given frequency were averaged. Indeed, similarly to what done here with total intensity flux densities, the PACO catalogue reports for each $2\,$GHz frequency band $4$ flux densities, one for each $512\,$MHz sub-band. Before performing any temporal average, we consider the median value over the $4$ chunks  to provide a value for each frequency. Then, the error associated to the averaged PACO flux densities $\sigma_{\rm PACO}$, is given by:
\begin{equation}
\sigma_{\rm PACO}=\sqrt{\sigma^2_{\rm max}+\sigma^2_{\langle S \rangle_{\rm PACO}}} \,,
\end{equation}
were $\sigma_{\rm max}$ is the maximum error over the four $512\,$MHz sub-bands and $\sigma_{\langle S \rangle_{\rm PACO}}$ the error associated to the average of the PACO fluxes over the selected epochs, $\langle S \rangle_{\rm PACO}$.

The AT20G data were collected at $4.86$, $8.64$ and $20\,$GHz with the old ATCA correlator set with a $2\times128\,$MHz contiguous bands for each frequency. We can straightforwardly compare these observations with ours at $5.5$, $9$ and $18\,$GHz, neglecting the small differences in the central frequencies.

Following Sadler et al. (2006), the variability index (V.I.) of a population is defined as:
\begin{equation}
\hbox{V.I.}=\frac{100}{\langle S \rangle}\sqrt{\frac{\displaystyle\sum_{i=1}^{n}\left( S_i-\langle S \rangle\right)^2-\displaystyle\sum_{i=1}^{n} \sigma_i^2}{n}},
\label{eq:VI}
\end{equation}
$\langle S \rangle$ being the average of the $n$ flux density measurements at a given frequency, $S_i$, having error $\sigma_i$.

In Table~\ref{tab:VarFreq} we report the mean V.I.'s in total intensity at each frequency for two time lags, $4-5$ and $8$ years, corresponding to the intervals between the present observations and the PACO or the AT20G ones, respectively. Variability indices were computed both for the full sample and for the steep- and peaked-spectrum populations. The errors provided in the table are the rms of the V.I.'s rescaled by the $\sqrt{N}$, where $N$ is the number of objects in the considered class.

For a time lag of $4-5$ years the mean V.I. of steep-spectrum sources increases with frequency, consistent with earlier  results (Impey \& Neugebauer 1988; Ciaramella et al. 2004, Bonavera et al. 2011). The increase is slightly milder for the full sample because it is not seen in the case of peaked-spectrum sources. No trend with frequency is found for the $8$ year lag.

\begin{figure}
\includegraphics[width=\columnwidth]{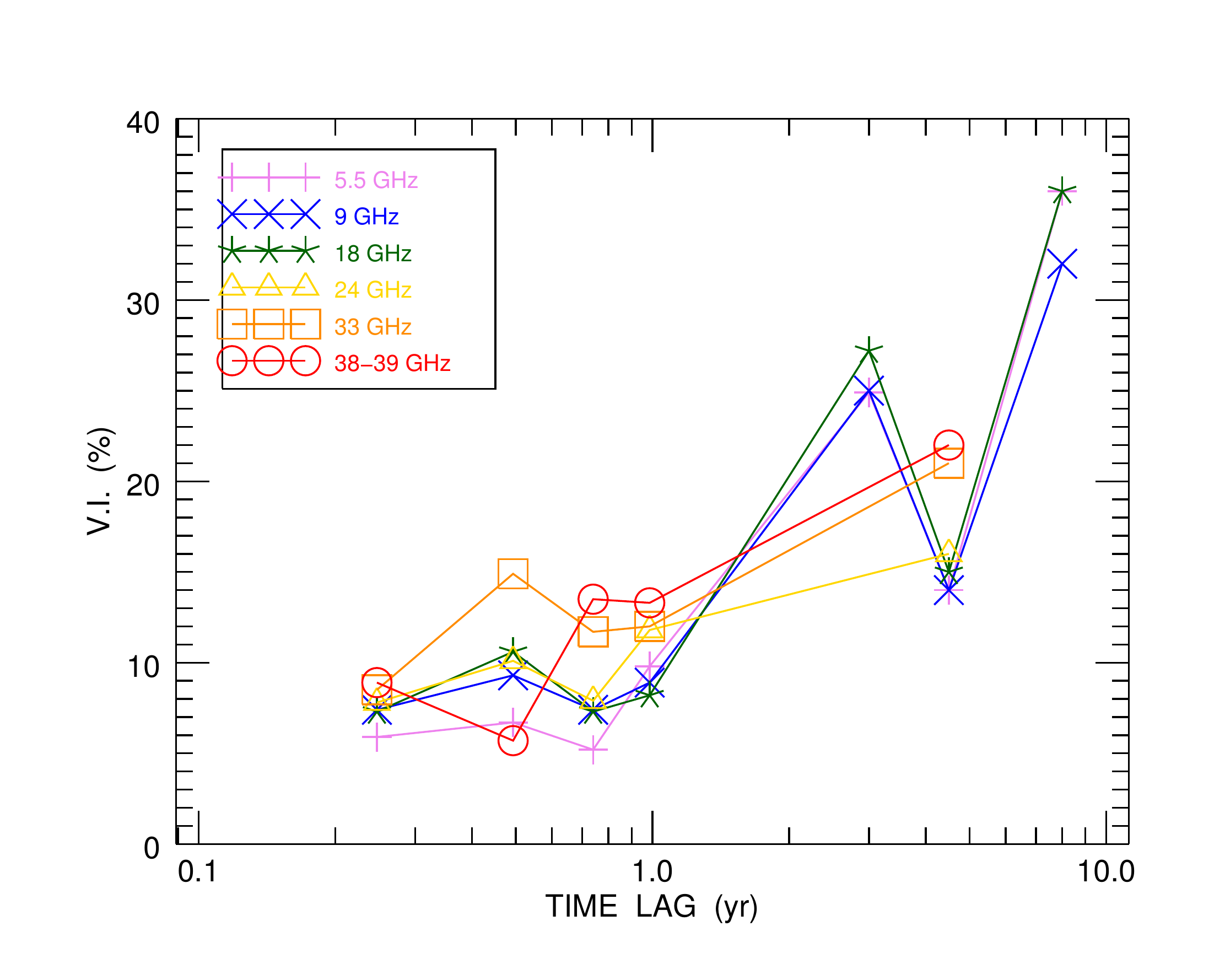}
\caption{Mean variability indices in total intensity {\it vs.} time lag at the observed frequencies (different colours). Variability measurements for the faint PACO sample are also displayed (Massardi et al. 2016).}
\label{fig:VIvstlag}
\end{figure}

Fig.~\ref{fig:VIvstlag} reports for each frequency (different colours) the variability index against the time lag. Our time lag coverage is complemented by variability measurements provided by Massardi et al. 2016 for the faint PACO sample: on average, there appears an increase of the variability index with the time lag at all the frequencies. Moreover, for those time lags for which there are also measurements at the highest frequencies, namely $33$ and $38\,$GHz, the variability indices are typically higher than those associated to lower frequencies.  

Due to the lack of PACO polarisation data we could estimate the V.I. only for the $8$ year lag (last line of Tab.~\ref{tab:VarFreq}). Also, due to the larger fractional uncertainties, we could make the analysis only for the full sample. The V.I. turned out to be substantially larger than in total intensity, with no significant frequency dependence.

\begin{table}
\caption{Mean variability indices in total intensity and in polarisation (last row).}
\label{tab:VarFreq}
\small\addtolength{\tabcolsep}{-4.3pt}
\begin{tabular}{c|c|c|c|c|c|c|c}
\hline
Sel.& Time (yr) & $5.5$ & $9$ & $18$ & $24$ & $33$ & $38$\\
\hline
All& $4-5$ & $14 \pm 2$ & $14 \pm 2$ & $15 \pm 2$ & $16 \pm 2$ & $21 \pm 2$ & $22 \pm 2$\\
& $8$ & $36 \pm 3$ & $32 \pm 3$ & $36 \pm 3$ & & & \\
Steep& $4-5$ & $10 \pm 2$ & $10 \pm 2$ & $12 \pm 2$ & $14 \pm 2$ & $21 \pm 3$ & $24 \pm 3$\\
& $8$ & $35 \pm 5$ & $32 \pm 5$ & $38 \pm 5$ & & & \\
Peaked & 4-5 & $20 \pm 3$ & $20 \pm 3$ & $18 \pm 3$ & $19 \pm 3$ & $20 \pm 4$ & $19 \pm 3$\\
& $8$ & $38 \pm 5$ & $32 \pm 5$ & $31 \pm 5$ & & & \\
\hline
All (pol.)& $8$ & $50 \pm 7$ & $57 \pm 6$ & $53 \pm 6$  & & & \\
\hline
\end{tabular}
\end{table}

\section{Conclusions}
\label{discusseconcl}

We have presented high sensitivity ($\sigma_P\simeq 0.6\,$mJy) polarimetric observations of a complete sample of $53$ compact extragalactic radio sources drawn from the ``faint'' PACO sample. Thanks to the high sensitivity we achieved a high ($\sim 91\%$) $5\,\sigma$ detection rate in polarisation.

The continuum spectra of over $95\%$ of sources are well fitted by double power laws, both in total intensity and in polarisation. However, substantial variations of the polarisation degree with frequency were found, implying different spectral shapes in polarisation compared to total intensity.

Within the spectral range covered by our observations ($5.5-38\,$GHz) most sources can be classified either as steep-spectrum or as peaked-spectrum, both in total intensity and in polarisation, although less than half of sources have the same classification in the two cases.

Between $28$ and $38\,$GHz all spectral indices are steep, so that the dichotomy between flat and steep spectra, associated to compact and extended sources, well established at low frequencies, no longer holds at these frequencies.

No significant trend of the fractional polarisation with either flux density or frequency was found, although we caution that the limited ranges covered by our sample and the variety of spectral behaviours may hamper the detection of weak trends. However, steep-spectrum sources show higher polarisation fractions than peaked-spectrum ones at all the observed frequencies.

We found evidence of Faraday rotation for $2$ sources. The derived values of the rotation measures are $(96\pm 13)\,{\rm rad}/{\rm m}^2$ and $(30\pm 13)\,{\rm rad}/{\rm m}^2$, respectively.

The mean variability index in total intensity of steep-spectrum sources increases with frequency for a $4-5$ year lag, while no significant trend shows up for peaked-spectrum sources and for the $8$ year lag. In polarisation, the variability index, that could be computed only for the $8$ year lag, is substantially higher than in total intensity and has no significant frequency dependence.

\section*{Acknowledgments}
We thank the anonymous referee for useful comments.
We acknowledge financial support by the Italian {\it Ministero dell'Istruzione, Universit\`a e Ricerca} through the grant {\it Progetti Premiali 2012-iALMA} (CUP C52I13000140001).
Partial support by ASI/INAF Agreement 2014-024-R.1 for the {\it Planck} LFI Activity of Phase E2 and by ASI through the contract I-022-11-0 LSPE is acknowledged.
We thank the staff at the Australia Telescope Compact Array site, Narrabri (NSW), for the valuable support they provide in running the telescope and in data reduction. The Australia Telescope Compact Array is part of the Australia Telescope which is funded by the Commonwealth of Australia for operation as a National Facility managed by CSIRO. AB acknowledges support from the European Research Council under the EC FP7 grant number 280127. VC acknowledges DustPedia, a collaborative focused research project supported by the European Union under the Seventh Framework Programme 
(2007-2013) call (proposal no. 606824). The participating institutions are: Cardiff University, UK; National Observatory of Athens, Greece; Ghent University, Belgium; Université Paris Sud, France; National Institute for Astrophysics, Italy and CEA (Paris), France. LT and LB acknowledges partial financial support from the Spanish Ministry of Economy and Competitiveness (MINECO), under project AYA-2015-65887-P.

\bsp


\end{document}